\documentclass[journal=mamobx,manuscript=article,psamsfonts]{achemso}

\usepackage{graphicx,pstricks,mathrsfs}
\usepackage{subfigure}
\usepackage{amsmath,color}    
\usepackage{hyperref,textcomp}   
\usepackage{eucal}
\usepackage{amssymb,amsfonts}
\usepackage[version=3]{mhchem}

\DeclareMathAlphabet{\mathcal}{OMS}{cmsy}{m}{n} 

\SectionNumbersOn
\title{Non-affine displacements in flexible polymer networks}

\author{Anindita Basu}
\email{abasu@sas.upenn.edu}
\affiliation{Department of Physics and Astronomy, University of Pennsylvania,PA 19104, USA}

\author{Qi Wen}
\affiliation{Institute for Medicine and Engineering, University of Pennsylvania, Philadelphia, PA 19104, USA}

\author{Xiaoming Mao}
\affiliation{Department of Physics and Astronomy, University of Pennsylvania,PA 19104, USA}

\author{T.~C.~Lubensky}
\affiliation{Department of Physics and Astronomy, University of Pennsylvania,PA 19104, USA}

\author{Paul A. Janmey}
\affiliation{Institute for Medicine and Engineering, University of Pennsylvania, Philadelphia, PA 19104, USA}
\alsoaffiliation{Department of Physics and Astronomy, University of Pennsylvania,PA 19104, USA}

\author{A. G. Yodh}
\affiliation{Department of Physics and Astronomy, University of Pennsylvania,PA 19104, USA}

\begin{document}

\begin{abstract}
The validity of the affine assumption in model flexible polymer networks is explored. To this end, the displacements of fluorescent tracer beads embedded in polyacrylamide gels are quantified by confocal microscopy under shear deformation, and the deviations of these displacements from affine responses are recorded. Non-affinity within the gels is quantified as a function of polymer chain density and cross-link concentration. Observations are in qualitative agreement with current theories of polymer network non-affinity. The measured degree of non-affinity in the polyacrylamide gels suggests the presence of structural inhomogeneities which likely result from heterogeneous reaction kinetics during gel preparation. In addition, the macroscopic elasticity of the polyacrylamide gels is confirmed to behave in accordance with standard models of flexible polymer network elasticity.
\end{abstract}

\maketitle
\section{Introduction}

Affine deformation is an essential assumption in the classical theory of elasticity. In the classical theory, deformation is assumed to be distributed homogeneously in the material so that strain is spatially constant at all length scales. The affine assumption permits elastic properties of cross-linked polymer networks to be readily derived from theories of rubber elasticity based on the entropy of a single polymer chain in the network. In practice, however, such affine deformations only occur in perfect crystals under very small deformation. In polymer networks, especially networks composed of semi-flexible or rigid filaments, the microscopic network deformations should be non-affine below certain length scale.

Non-affinity can arise from different sources. In near-ideal flexible polymer melts, deformations might be expected to be affine on length scales much larger than the average mesh size and non-affine at lengths scales of the order of the mesh size or smaller~\cite{1}. Random thermal fluctuations of the cross-link junctions, along with thermal undulations of the polymer chains may also lead to non-affine behavior in polymer gels. Inhomogeneities introduced into the network micro-structure during sample preparation can also introduce non-affine responses; such inhomogeneities might be expected to be a function of reaction kinetics and other sample preparation parameters~\cite{2}.

Over the years, the connection between shear deformation and non-affinity has been explored theoretically in a wide range of materials including rubber-like spatially homogeneous elastic media~\cite{2}, entangled or cross-linked polymer networks~\cite{3, 4, 5, 6, 7}, semi-flexible polymer networks with rigid~\cite{8} and flexible cross-links~\cite{9}, stiff rod networks~\cite{10}, biopolymers~\cite{11}, amorphous systems~\cite{12} and foams~\cite{13}. Indeed, it has been proposed that non-linear elasticity in polymer networks has its origin in non-affine responses~\cite{1}. In spite of continued interest in this problem and its fundamental importance, relatively little experimental quantification of the non-affine phenomenon has been carried out in semi-flexible biopolymer gels~\cite{15,20}, and we are not aware of any non-affinity studies for the simple case of flexible polymers. Experiments along these lines should provide benchmarks for future understanding of the subject.

This paper describes an investigation of non-affine shear deformations in a model flexible polymer gel: polyacrylamide gels with bisacrylamide cross-links. Polyacrylamide is well suited for the investigation because it is comparatively well-controlled, and its stiffness is tunable by the number of bisacrylamide cross-links. As part of this study, macroscopic rheological measurements are carried out to confirm the simple rubber-like elastic character of these networks.  Then deformation fields in the gels under external shear stress are characterized by measuring the displacements of fluorescent beads entrapped in the gels. Bulk rheology and confocal microscopy are used in concert for the latter study. A non-affine parameter, $\mathcal{A}$, is defined to quantify the degree of non-affinity in the displacement field. $\mathcal{A}$ is measured as a function of bead size, polymer chain density, and cross-link density in the gels.  We test simple predictions of a recently developed theory of non-affinity in random elastic media~\cite{2} and obtain estimates for the fluctuations in elastic modulus of the gels from $\mathcal{A}$.

\section{Experiment}

\subsection{Sample Preparation}

The polyacrylamide (PA) gel is prepared by polymerizing acrylamide monomers and bisacrylamide (bis) cross-links in aqueous 50~mM HEPES buffer at pH $=8.2$, using free-radical polymerization reaction initiated by 0.1\% weight/weight (w/w) ammonium persulphate (APS) and 0.3\% w/w N, N, N', N'- tetramethylethylenediamine (TEMED). (Here the percent of X w/w equals the mass in grams of X per 100 grams of solution.)   Fluorescent polystyrene tracer beads are mixed into our solution at a concentration of 0.004\% weight per volume (w/v), before the addition of bis cross-links. (Here the percent of X w/v equals the mass in grams of X dissolved/suspended in 100 milliliters of solvent.) Thus, a tracer bead concentration of 0.004\% w/v is attained by dissolving 0.004 gram of tracer beads in 100 milliliters of water. This procedure helps to distribute the beads uniformly throughout the polymer network. Internally labeled and carboxylate-modified fluorescent polystyrene micro-spheres of various diameters are used for this purpose, viz., 0.6 $\mu \mbox{m}$, 1 $\mu \mbox{m}$ (Molecular Probes, California, USA), and 1.5 $\mu \mbox{m}$ (Bangs laboratories Inc., Indiana, USA). Acrylamide (7.5\%, 15\% w/v and bisacrylamide (0.03 - 0.12\%w/v) concentrations are systematically changed to study the effects of polymer concentration, cross-link density and mesh size on the polymer network rheology.

\subsection{Rheology}

Rheology measurements are performed using a stress-controlled Bohlin Gemini rheometer (Malvern Instruments, UK), with a cone and plate geometry of $4^{\circ}$ cone angle and 20 mm diameter. Samples are prepared \textit{in situ} so that good contact is routinely established between the sample surfaces and the rheometer plates to prevent slippage at high strains. The shear modulus ($G'$) and loss modulus ($G''$) for each sample during the process of polymerization are monitored using low strain amplitude ($\gamma_0$= 0.01) and low frequency ($f=0.1$ Hz) oscillatory shear measurements. The polymerization reaction proceeds for $\sim 30$ minutes, with the elastic and viscous moduli attaining steady-state values in less than $10$ minutes. Care is taken to prevent solvent evaporation by sealing off the sample from the sides with a low density, low viscosity ($\sim 50~$mPa$\cdot$s) silicone oil. The elastic and viscous moduli, $G'$ and $G''$, respectively, for these gels are measured as functions of frequency, amplitude and temperature.  These measurements are intended to confirm that the gels behave in accordance with the existing theories of flexible polymer networks~\cite{16}. A set of control experiments are performed on the PA gels, with and without the tracer beads, to further confirm that macroscopic properties of the gels are not altered by the addition of the tracer beads.

\begin{figure}[htp]
\subfigure[]{\includegraphics[trim = 5mm 45mm 5mm 15mm, clip, width=0.8\textwidth]{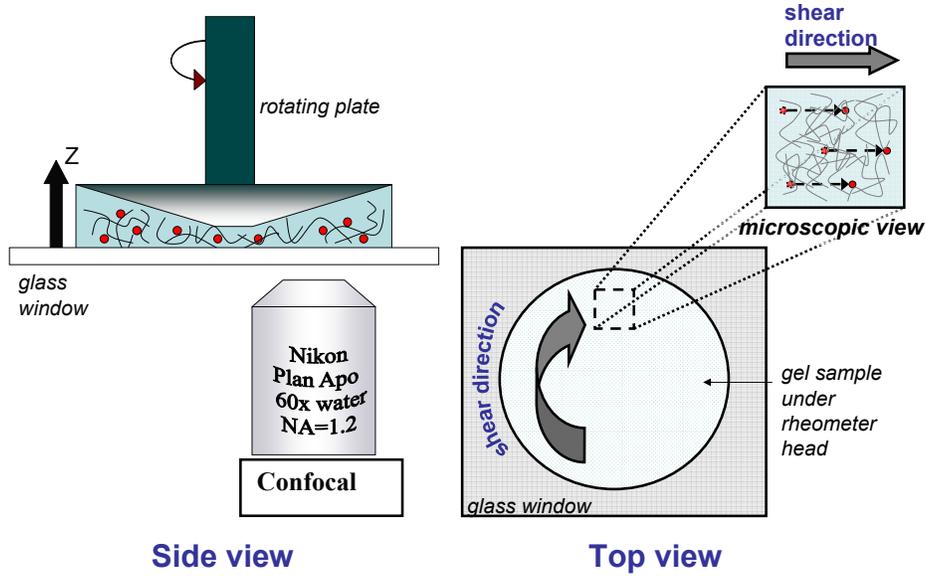}}\\
\subfigure[]{\includegraphics[trim = 35mm 30mm 35mm 23mm, clip, width=0.45\textwidth]{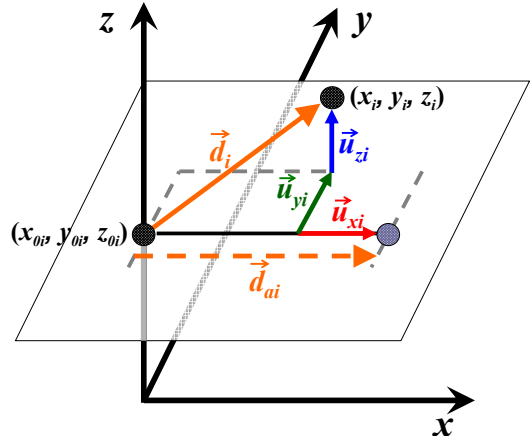}}\\
\caption{\label{fig:fig1}Experimental Setup. (a) Experimental schematic. (b) Sketch of the non-affine displacements of tracer beads. $(x_{0i},y_{0i},z_{0i})$ and $(x_{i},y_{i},z_{i})$ mark the positions of a tracer bead without and under shear, respectively. Dashed arrow indicates affine displacement, $\vec{d}_{ai}$, of tracer bead in the direction of shear (x-axis).  $\vec{d}_{i}$ is the measured displacement of the tracer bead. $\vec{u}_{xi}$, $\vec{u}_{yi}$, $\vec{u}_{zi}$ indicate the non-affine deviations along the $x$, $y$ and $z$ axes respectively. $\vec{u}_{i}=\vec{d}_{ai}-\vec{d}_{i}$ is the non-affine deviation. }
\end{figure}

\subsection{Confocal Microscopy}

Microscopic deformation of the PA gels under shear is studied by tracking tracer bead displacements in the sample using confocal microscopy. A VTeye confocal system (VisiTech International, UK) is used in conjunction with an inverted Eclipse TE200 microscope (Nikon Instruments, USA) for this purpose. The lower plate of the rheometer is replaced by a home-built transparent sample holder and is mounted on the microscope to permit visualization of the samples under shear [Fig.~\ref{fig:fig1}(a)].

A $60\times$ water objective (NA = 1.2) is used to visualize the sample over a depth of 100 $\mu$m. 3D stacks map the entrapped tracer beads $(70~\mu \mbox{m}\times70~\mu \mbox{m}\times60~\mu \mbox{m})$ using the confocal setup with and without applied shear. The step size of the 3D stacks is varied from 100 nm to 200 nm for tracer bead sizes ranging from 600 nm to 1.5 $\mu$m. A wide range of shear strain, up to 50\% amplitude, is applied. 

The image stacks are processed using relatively standard Matlab routines, which determine the beads' positions with subpixel accuracy~\cite{17},~\cite{18}. For each stress value, a set of two image stacks are taken, one with shear and one without. The 3D locations of beads in a gel without external shear stress are determined as $(x_{0i}, y_{0i}, z_{0i}) $, for $i=1,2,\ldots,N$, where $N$ is the number of tracked beads. The centroids of the corresponding $N$ beads in the image stack under shear stress are measured too, as $(x_i, y_i, z_i)$; for convenience the direction of shear is taken to be along the $\vec{x}$ axis in the figure. The displacements of tracer beads are then calculated from the tracking results as $\vec{d}_{i}=(x_{i}-x_{0i}, y_{i}-y_{0i}, z_{i}-z_{0i})$. The system permits the displacements of tracer beads to be measured with a spatial resolution of 50 nm. On average, 30 beads are tracked in each 3D stack. 

\begin{figure}[hbp]
\subfigure[]{\includegraphics[trim = 31mm 70mm 40mm 70mm, clip, width=0.45\textwidth]{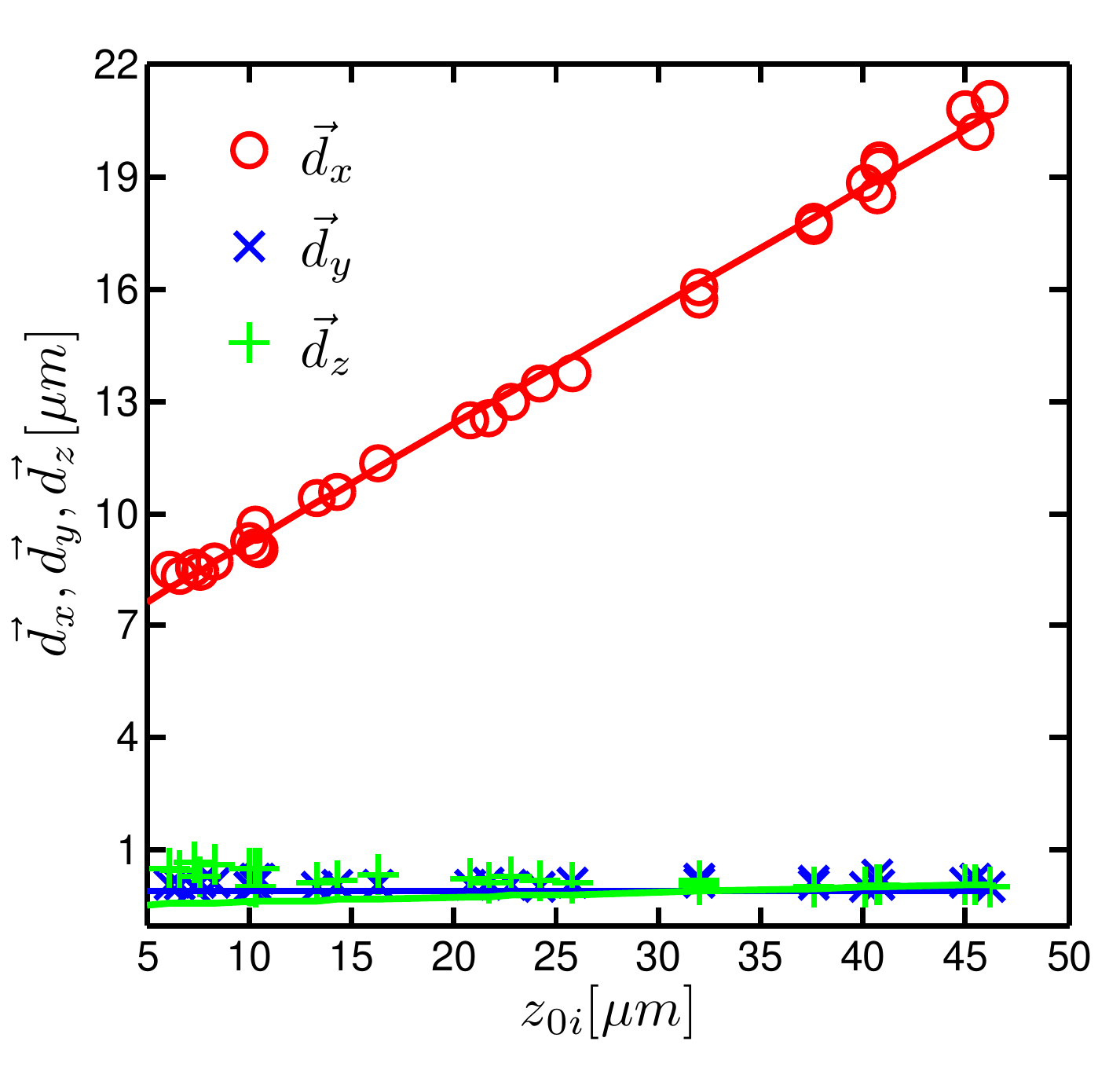}}
\subfigure[]{\includegraphics[trim = 7mm 5mm 20mm 15mm, clip, width=0.4\textwidth]{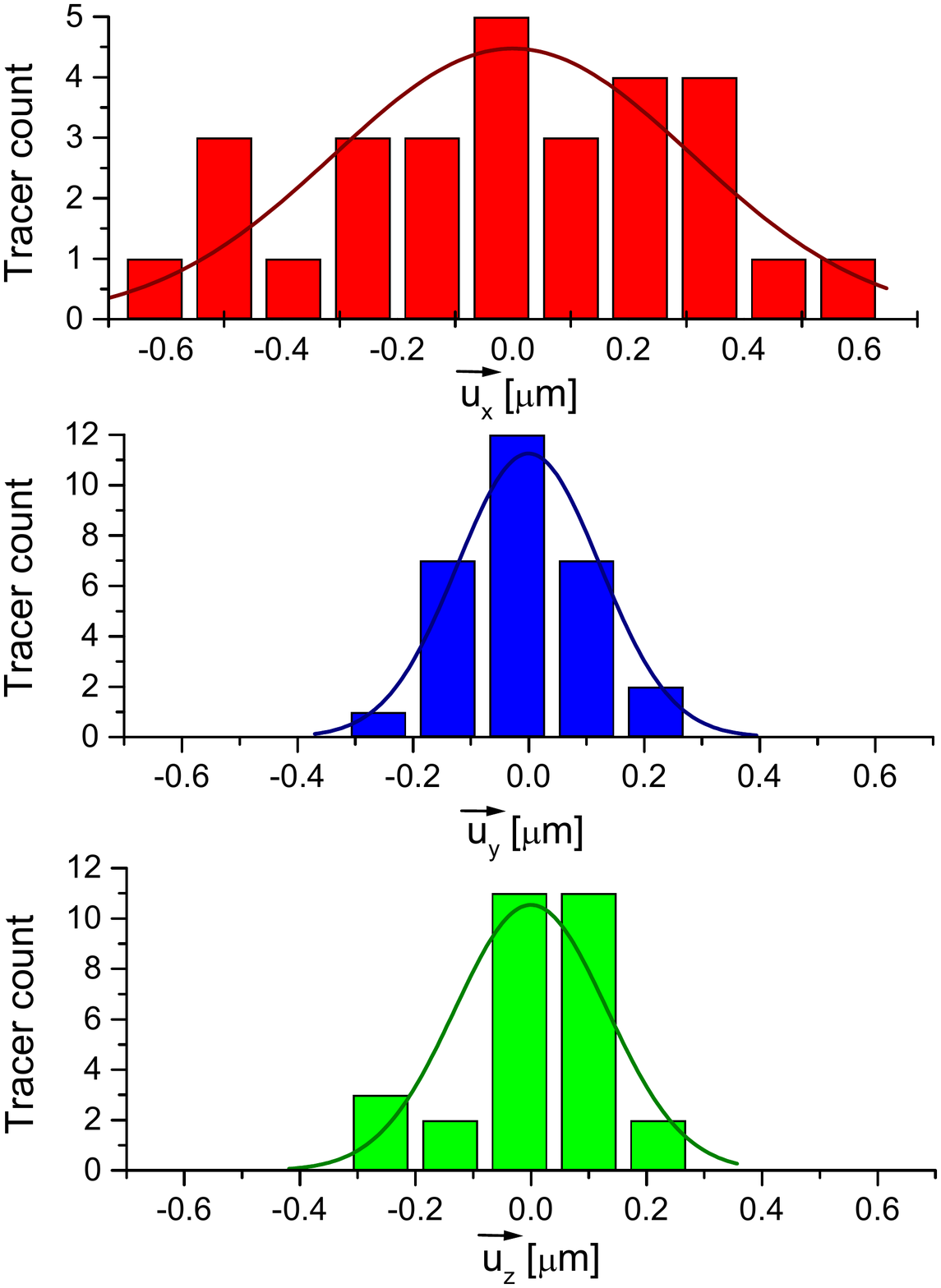}}
\caption{\label{fig:fig1b} (a) Experimentally measured displacements of tracer beads in the direction of shear, $\vec{d}_{i}$, that has been decomposed along x-(red circles), y-(blue crosses) and z- (green pluses) axes, as a function of the distance, $z_{0i}$ from the fixed lower plate of the rheometer. This sample is 7.5\% acrylamide and 0.03\% bis PA gel. The solid lines give the strains, $\gamma_x$, $\gamma_y$, and $\gamma_z$ obtained from their linear fits. Note that $\gamma_y$, and $\gamma_z$ are $\approx 0$.  (b) The distribution of non-affine deviations of tracer beads for the same sample PA gel shown in Fig.~\ref{fig:fig1b}(a), at $\gamma~=0.3$, decomposed along the x-, y- and z- axes. The measurements are normally distributed around the \textit{affine} displacement position, as indicated by the solid curves. }
\end{figure}

\subsection{Non-affine Parameter}
A measure of the degree of non-affinity is provided by the non-affine parameter $\mathcal{A}$, which is defined in Ref.~\cite{2} as:
\begin{eqnarray}
\mathcal{A}=\frac{1}{N}\sum^{N}_{i=1}| \vec{u}_i |^2.\nonumber
\end{eqnarray}
Here $\vec{u}_i=\vec{d}_{i}-\vec{d}_{ai}$, is the deviation of the measured tracer-bead displacement, $\vec{d}_{i}$, from the affine displacement, $\vec{d}_{ai}$ [Fig.~\ref{fig:fig1}(b)]. 

For a perfect shear deformation along the $x$-axis, the affine displacement $\vec{d}_{ai}$ would be in the direction of shear only-- the $y$ and $z$ components must be zero. We measure the resultant strains along all three component axes, $\gamma_x$, $\gamma_y$ and $\gamma_z$ by fitting the $x$, $y$ and $z$ components of $d_{i}$ to linear functions of $z_{0i}$, as seen from a sample PA gel (7.5\% acrylamide, and 0.03\% bis) under an applied strain of $\gamma~=0.3$ in Fig.~\ref{fig:fig1b}(a). The real strain on the sample is determined as $\gamma=\sqrt{\gamma_x^2+\gamma_y^2+\gamma_z^2}$. The $x$, $y$ and $z$ components of the affine displacement vector, $\vec{d}_{ai}$ are then calculated as $z_{0i}\gamma_x$, $z_{0i}\gamma_y$, and $z_{0i}\gamma_z$. Note that the $y$ and $z$ components, both perpendicular to the direction of shear, do not vary as a function of $z_i$, resulting in $\gamma_y$ and $\gamma_z \approx 0$ . Fig.~\ref{fig:fig1b}(b) plots the distribution of non-affine deviations, $\vec{u}_x$, $\vec{u}_y$, and $\vec{u}_z$ for the same sample gel, along the $x$, $y$, $z$ axes respectively, for the same strain of $\gamma~=0.3$ as seen in Fig.~\ref{fig:fig1b}(a). $|\vec{u}_i|^2$ is calculated as $(x_i-x_{0i}-\gamma_xz_{0i})^2+(y_i-y_{0i}-\gamma_yz_{0i})^2+(z_i-z_{0i}-\gamma_zz_{0i})^2$. 

The non-affine parameter, $\mathcal{A}$ is then defined in terms of these variables as
\begin{eqnarray}
\mathcal{A}=\frac{1}{N}\sum^{N}_{i=1}[(x_i-x_{0i}-\gamma_xz_{0i})^2+(y_i-y_{0i}-\gamma_yz_{0i})^2+(z_i-z_{0i}-\gamma_zz_{0i})^2].\label{eq:eq1}
\end{eqnarray}

\section{Results}
 
\subsection{Bulk Rheology Measurements}

\begin{figure}[h!]
\subfigure[]{\includegraphics[trim = 20mm 20mm 40mm 30mm, clip, width=0.40\textwidth]{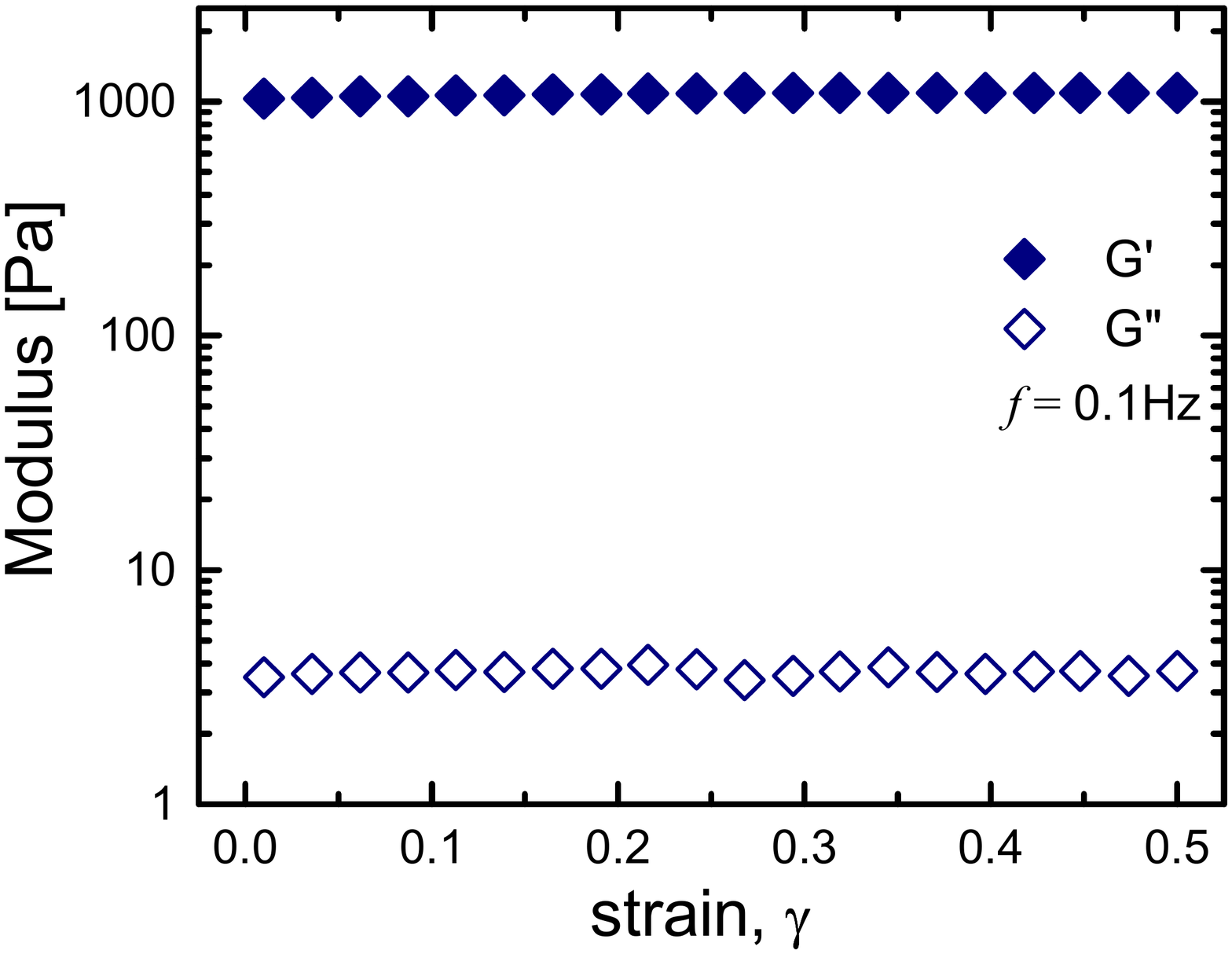}}
\subfigure[]{\includegraphics[trim = 20mm 25mm 20mm 10mm, clip, width=0.45\textwidth]{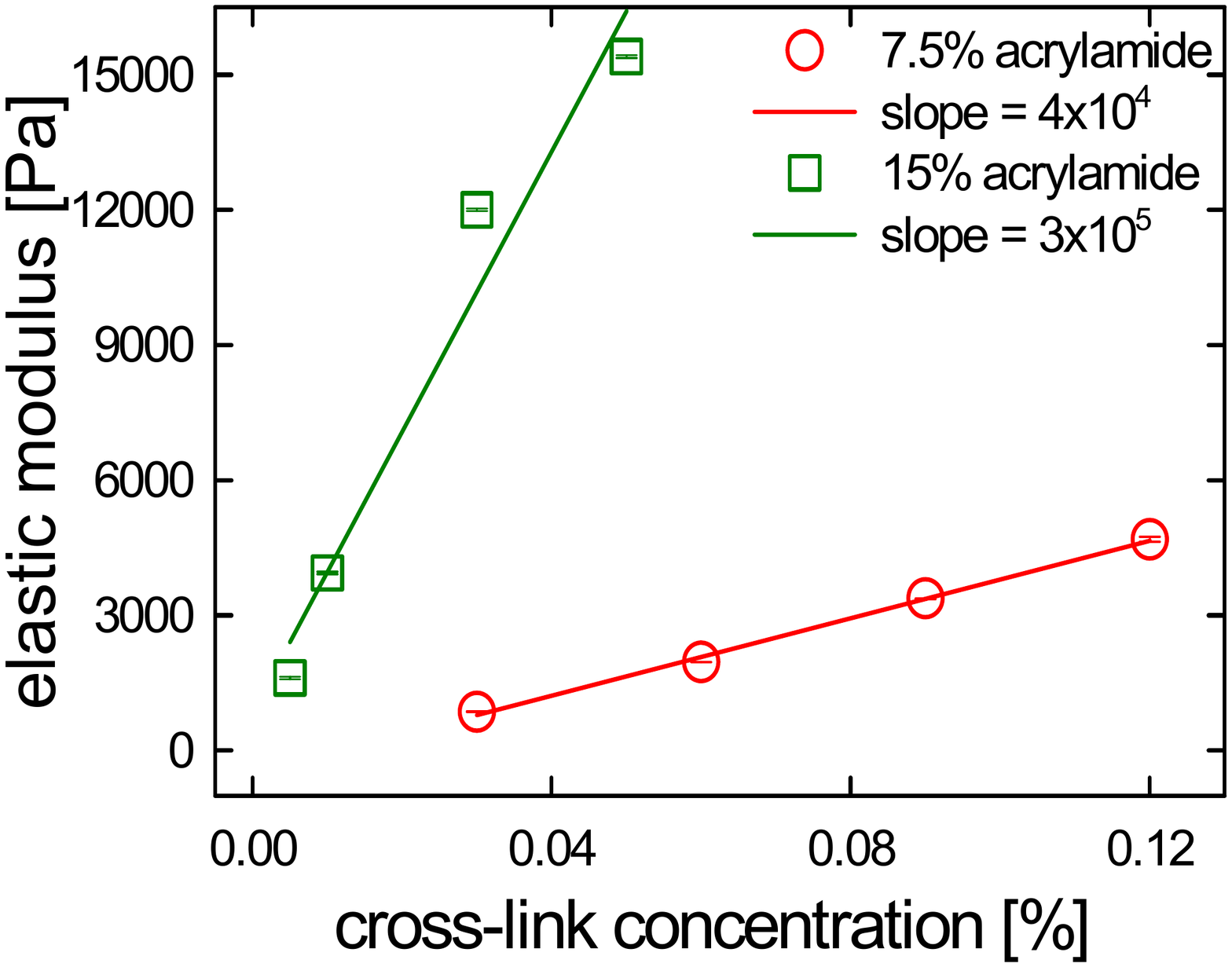}}
\subfigure[]{\includegraphics[trim = 33mm 25mm 35mm 30mm, clip, width=0.45\textwidth]{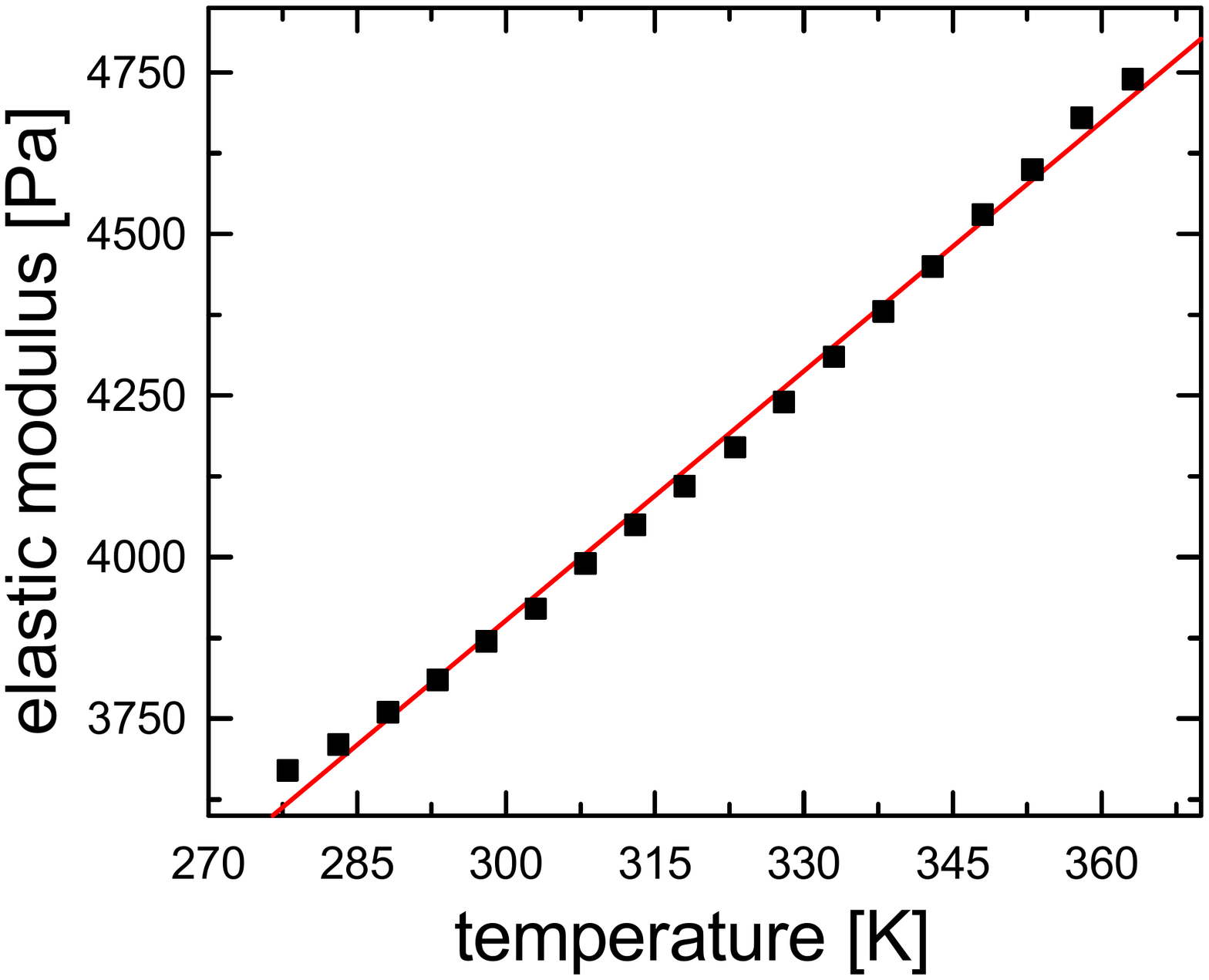}}
\caption{\label{fig:fig2} Rheology of polyacrylamide gels. (a) $G'$ of sample PA gel is two orders of magnitude larger than the $G"$ and theses values remain constant over a wide range of applied strain. Data are shown for a gel with 7.5\% acrylamide at 0.03\% bisacrylamide cross-link concentration, at an oscillatory frequency of 0.1 Hz. (b) $G'$ of 7.5\% and 15\% polyacrylamide gels as a function of cross-link concentrations. Error bars denote standard deviations which are less than 2\% of the mean elastic moduli. The solid lines indicate linear fits to the data. Note that the overall moduli of the gels with 7.5\% acrylamide are significantly lower than that of 15\% acrylamide for comparable cross-link density. (c) $G'$ as a function of temperature (red line is the linear fit). Data are shown for a PA gel with 7.5\% acrylamide with 0.09\% bisacrylamide.}
\end{figure}

The PA gels used in our experiments are solid-like materials with $G'$ ranging from $7.7\times10^2$ Pa to $1.5\times10^4$ Pa, 2 to 3 orders of magnitude larger than the $G''$. In Fig.~\ref{fig:fig2}(a), $G'$ and $G''$ of a gel made of 7.5\% acrylamide and 0.06\% bisacrylamide are plotted as functions of the amplitude of the oscillatory shear strain at oscillation frequency, $f=0.1$ Hz. $G'$ is approximately 100 times larger than $G''$. Moreover, both $G'$ and $G''$ are independent of the applied shear strain for strains up $\gamma=0.5$, confirming the linear elastic response of PA gels. The frequency response of PA gels is characterized by measuring $G'$ and $G''$ at oscillatory strains with amplitude $\gamma_0=0.01$ and frequency ranging from 0.1 Hz to 100 Hz. Within this frequency range, $G'$ remains constant, and $G''$ increases with increasing frequency (data not shown).

The elastic moduli of our PA gels vary linearly with bisacrylamide concentration and sample temperature. Cross-link and monomer concentration trends are shown in Fig.~\ref{fig:fig2}(b).  Notice, when the bisacrylamide concentration increases from 0.03\% to  0.12\%, $G'$ for gels with 7.5\% acrylamide increases linearly from $7.7\times10^2$ Pa to $4.9\times10^3$ Pa. Similarly, $G'$ for 15\% acrylamide PA gels increases from $1.6\times10^3$ Pa at 0.005\% bis to $1.5\times10^4$ Pa at 0.05\% bis concentration. We also investigated the temperature dependence of the network elasticity within the attainable temperature range of the rheometer, i.e. $5^\circ\mbox{C} <T< 90^\circ\mbox{C}$. In Fig.~\ref{fig:fig2}(c), we show that $G'$ from the gel made of 7.5\% acrylamide and 0.09\% bisacrylamide increases linearly with sample temperature. This linear dependence of $G'$ on cross-link concentration and sample temperature follows the predictions of classical rubber elasticity theory. Note that the slopes of the linear fits of $G'$ as a function of cross-link concentrations for the 7.5\% acrylamide is lower than that of 15\% acrylamide PA gels. We suggest that some bis molecules form efficient cross-links and others do not, and that this difference in the slopes of G' versus cross-link concentrations for 7.5\% and 15\% acrylamide is due to the difference in effectiveness of the bis molecules in forming efficient cross-links, which increases with increasing monomer concentration. We discuss these effects further in Section \ref{CROSSLINKS}.

\subsection{Non-affine parameter, $\mathcal{A}$ scales as the square of the applied strain}
Confocal microscopy is used to visualize and record the displacements of the fluorescent tracer beads entrapped within a $(70~\mu \mbox{m}\times70~\mu \mbox{m}\times 60~\mu \mbox{m})$ volume in the PA gel. Since the tracer beads' size of $\sim 1~\mu m$ is much larger than the average mesh size of the PA gel, free Brownian motion is suppressed.  Within this small volume, located approximately 1 cm from the axis of rotation, the macroscopic shear strain applied to the beads can be approximated to be unidirectional. 

\begin{figure}[h!]
\subfigure[]{\includegraphics[trim = 27mm 75mm 30mm 75mm, clip, width=0.48\textwidth]{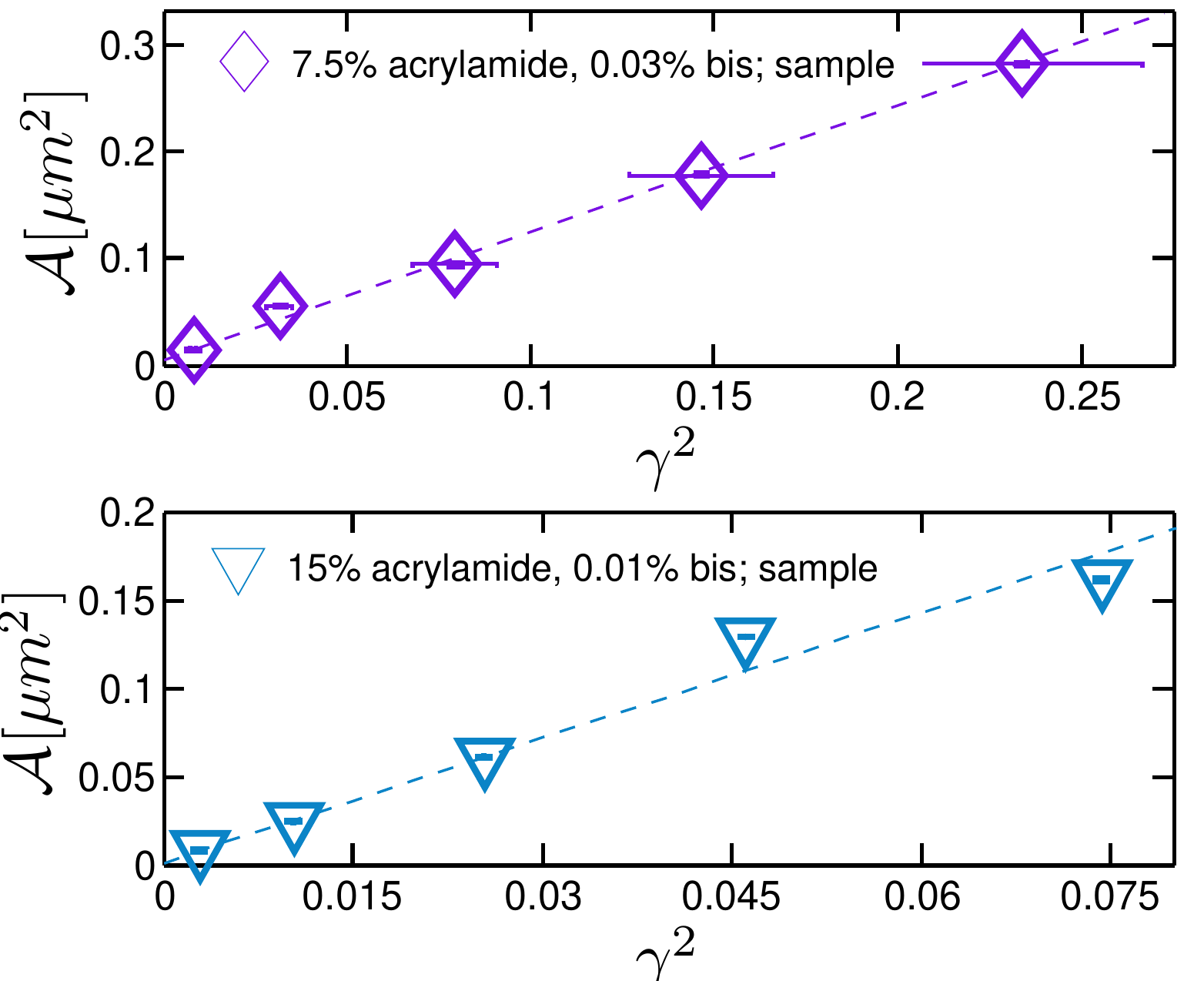}}
\subfigure[]{\includegraphics[trim = 35mm 75mm 35mm 75mm, clip, width=0.48\textwidth]{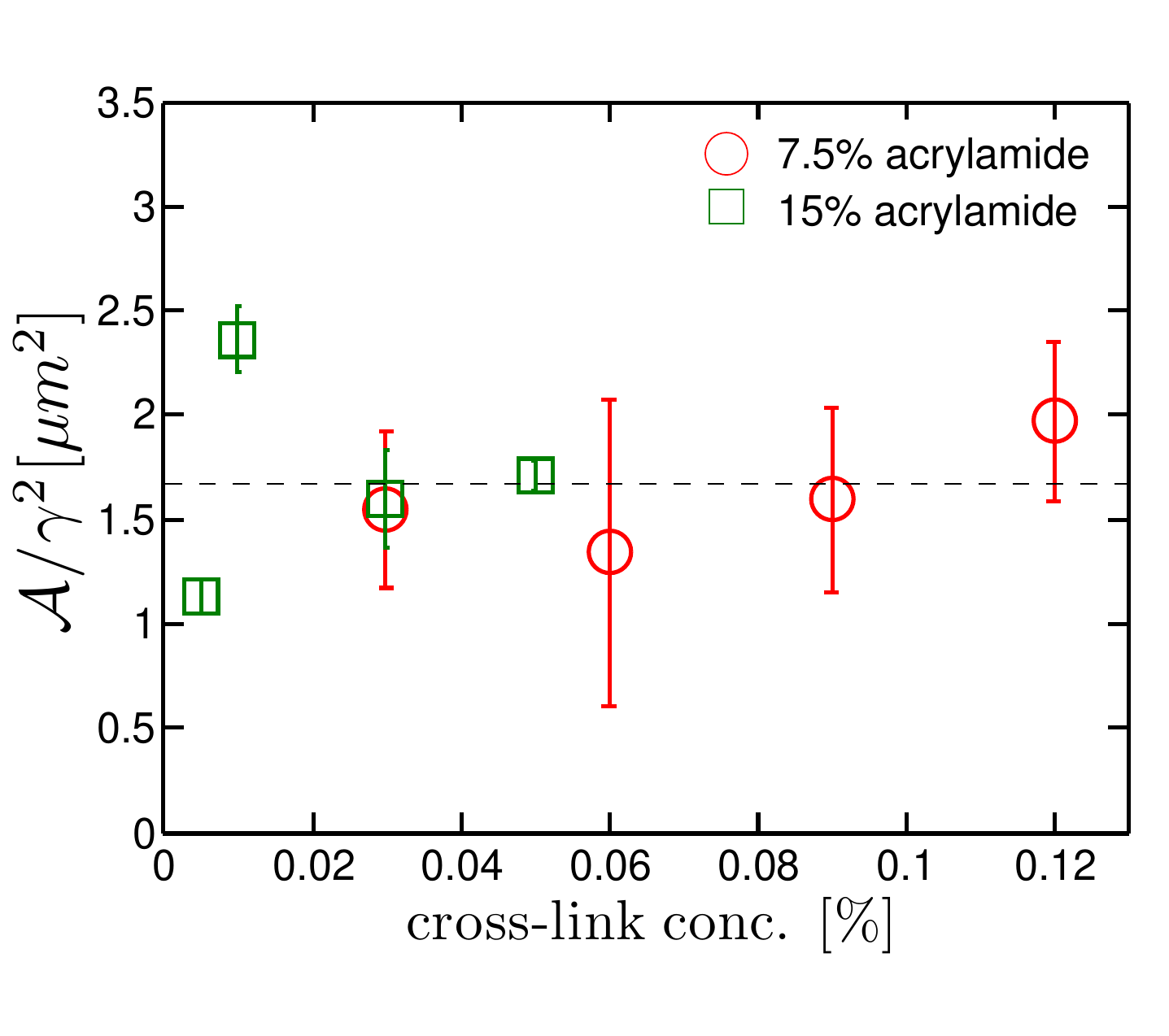}}
\caption{\label{fig:fig3} (a) The non-affine parameter scales as the square of the external strain field, as seen for sample polyacrylamide gels at 7.5\% acrylamide and  0.03\% bis (top), and 15\% acrylamide and  0.01\% bis (bottom). The dashed lines give best linear fits to the data. Error bars represent standard deviation of measurements of strain and non-affinity, the latter being smaller than symbol size. (b) Strain-normalized non-affine parameter, $\frac{\mathcal{A}}{\gamma^2}$ for sample PA gels at 7.5\% and 15\% acrylamide are plotted at varying bis concentrations. The data points and error bars represent the average value and standard error of measurements from different samples prepared in the same manner. The dashed line indicates the average $\frac{\mathcal{A}}{\gamma^2}$ calculated from all data points in the figure. }
\end{figure}  

In Fig.~\ref{fig:fig1b}(a), bead displacements along the $x$, $y$, and $z$ axis,  are plotted as a function of $z_{0i}$, the distance between the beads and the bottom surface. The displacements along the direction of shear, \textit{viz.}, the x-axis, increase linearly with $z_{0i}$, as expected from macroscopic shear deformation. Fitting $d_x$ to a linear function of $z_{0i}$ yields the strain $\gamma_x \approx \gamma$. $d_y$ and $d_z$, both perpendicular to the shear direction, are independent of $z_{0i}$ as shown in Fig.~\ref{fig:fig1b}(a). Also notice from Fig.~\ref{fig:fig1b}(b) that the non-affine displacements along each axis, \textit{viz.}, $\vec{u}_x$, $\vec{u}_y$ and $\vec{u}_z$, are much larger than the resolution of our system in the $xy-$ plane ($\sim 50~nm$) and comparable to that along the z axis ($\sim 80~nm$) and are normally distributed with mean value zero, i.e., distributed around the affine displacement positions. These uncertainties in tracking lead to a noise floor in $\mathcal{A}\sim 0.007 \mu m^2$.

The non-affine parameter $\mathcal{A}$ is readily computed from the measured bead displacements using Eq.~(\ref{eq:eq1}) for PA gels (7.5\% and 15\% acrylamide and a range of bisacrylamide concentrations). In Fig.~\ref{fig:fig3}(a), $\mathcal{A}$ increases with applied strain $\gamma$, and clearly scales as $\gamma^2$. 

In Ref.~\cite{2}, DiDonna and Lubensky developed a perturbation theory for non-affine deformations in solids with random, spatially inhomogeneous elastic moduli.  They characterized the non-affine deformations using the non-affinity correlation function
\begin{eqnarray}
	\mathcal{G}_{ij}(x,x')=\langle \vec{u}_{i}(x) \vec{u}_{j}(x') \rangle ,
\end{eqnarray}
where $u(x)$ is the nonaffine displacement field, and $\langle\cdot\rangle$ represents the average over randomness in the elastic moduli (i.e., a disorder average).  Because the disorder averaged quantities are translationally and rotationally invariant in the gel we consider, the correlation function only depends on the distance $\vert x-x'\vert$, and thus it is characterized by the Fourier transform $\mathcal{G}(q)\equiv \int d(x-x') \mathcal{G}_{ii}(x,x') e^{-i q\cdot (x-x')}$.  In Ref.~\cite{2} it is proved that this correlation function is related to the correlation function of the inhomogeneous elastic modulus $K$ as
\begin{eqnarray}\label{EQ:Gq}
	\mathcal{G}(q)\sim \frac{\gamma^2 \Delta^{K}(q)}{q^2 K^2}
\end{eqnarray}

where $K$ is the disorder averaged elastic modulus, and $\Delta^{K}(q)$ is the Fourier transform of the spatial correlation function of the elastic modulus $K$.  

In this theory, the zeroth order problem concerns elastic deformations in a homogeneous media of elastic modulus $K$, and the randomness in $K$ is treated as a perturbation from this homogeneous state.  To first order, the driving forces of the non-affine deformations are thus proportional to the zeroth order deformations, which are proportional to $\gamma$.  Therefore, to first order in perturbation theory, $\mathcal{G}(q)$ is proportional to $\gamma^2$.

The non-affine parameter $\mathcal{A}$ defined in the present experiment corresponds to $\mathcal{G}_{ii}(x,x)$,  
\begin{eqnarray}\label{EQ:AInt}
	\mathcal{A} = \mathcal{G}_{ii}(x,x) = \int\frac{d^3 q}{(2\pi)^3} \mathcal{G}(q)
	\sim \int\frac{d^3 q}{(2\pi)^3} \frac{\gamma^2 \Delta^{K}(q)}{q^2 K^2}.
\end{eqnarray}
It is clear from this equation that $\mathcal{A}\propto\gamma^2$.  In our experiment the relation $\mathcal{A}\propto\gamma^2$ is verified, as shown in Fig.~\ref{fig:fig3}(a).  This fairly robust relation has also been found in non-affine correlation functions of, for example, flexible polymer networks~\cite{19} and semiflexible polymer networks at small strain~\cite{20,21}.

The quantity $\frac{\mathcal{A}}{\gamma^2}$, which is independent of strain $\gamma$, provides a good measure of the degree of non-affinity of the sample. We shall refer to this quantity, $\frac{\mathcal{A}}{\gamma^2}$, as the strain-normalized non-affine parameter. $\frac{\mathcal{A}}{\gamma^2}$ is calculated for each sample as follows: For a particular strain, $\mathcal{A}$ is calculated by averaging the square of the non-affine displacements, ${\vec{u}_i}^2$, for all tracer beads in the sample; typically we carried out multiple shear measurements at the same strain (see Section \ref{REPEATED}), and the displacement data from all particles in all repeated shear measurements were averaged together to derive the mean $\mathcal{A}$ and its standard deviation. The resultant $\frac{\mathcal{A}}{\gamma^2}$ data were then  fit to a linear function. The slope of the linear fit gives $\frac{\mathcal{A}}{\gamma^2}$ for the sample; the intercept form the fitting is comparable to the noise floor of the measurements in $\mathcal{A}$. Standard deviations for the slopes were also derived. We use this parameter, $\mathcal{A}$ versus $\gamma$, which represents an intrinsic material property, for comparisons among samples prepared at different times or under different conditions. 

The $\frac{\mathcal{A}}{\gamma^2}$ value calculated for each sample, along with its constituent acrylamide and bis concentration is listed in Table 2 in the supporting material available online.  Fig.~\ref{fig:fig3}(b) plots the mean and standard error of the strain-normalized non-affinity parameter, $\frac{\mathcal{A}}{\gamma^2}$ for PA gel samples at various monomer (\textit{viz.,} 7.5\% and 15\% acrylamide, w/v), and cross-link (between 0.005\% and 0.12\% bisacrylamide, w/v) concentrations. The large error bars in the $\frac{\mathcal{A}}{\gamma^2}$ values for the 7.5\% acrylamide samples at different bis concentration (standard deviation $\sim 38\%$) arises primarily from sample-to-sample variations associated with gels polymerized under (ostensibly) identical experimental conditions.  The error bars for the 15\% acrylamide samples are much smaller than the 7.5\% samples, because data in the former case were extracted from a single sample polymerized at the given acrylamide and bis concentration.  Within this relatively large range of values, the strain-normalized non-affinity measure does not appear to vary significantly as a function of either the density of polymer chains or the network mesh-size, and is evenly distributed around the mean $\frac{\mathcal{A}}{\gamma^2}$ calculated over the entire range of PA gels sampled in our experiments. This mean value is indicated by the dashed line in Fig.~\ref{fig:fig3}(b).  The average values of $\frac{\mathcal{A}}{\gamma^2}$ obtained for different monomer concentrations along with their respective standard deviation and standard error of each group of measurements are summarized in Table~\ref{TAB:Summary}. We also used an alternative approach for calculating the mean $\frac{\mathcal{A}}{\gamma^2}$ in Section 4 of the supporting material available online; this alternative approach treated all beads across all samples equally. The results obtained by this alternative method were essentially same as the results above.  Perhaps not surprisingly, we will propose below that this measured non-affinity is largely dominated by inhomogeneities formed during synthesis of the PA gels, rather than being dictated by the thermal fluctuations of the polymer cross-links. 

\begin{table}[h!]
\begin{tabular}{|c c c c|}
\hline
monomer conc. &  $\frac{\mathcal{A}}{\gamma^2}$  &  Std. Dev. &  Std. Err.\\
\hline 
7.5\% & 1.65 & $\pm~$0.63 & $\pm~$0.21 \\ 
15\%  &  1.70  & $\pm~$0.51 & $\pm~$0.25\\ 
all samples & 1.67  & $\pm~$0.57 & $\pm~$0.16 \\ 
\hline
\end{tabular}
\caption{Summary of $\frac{\mathcal{A}}{\gamma^2}$ at different acrylamide concentrations.}
\label{TAB:Summary}
\end{table}

\section{Discussion}
\subsection{Effectiveness of cross-links}
\label{CROSSLINKS}
The measured $G'$ of polyacrylamide gels generally follows predictions of standard theories of rubber elasticity, i.e.,  $G'=2\nu Nk_BT$ where $N$ is the number density of cross-links, $k_B$ represents the Boltzmann constant, $T$ represents temperature, and $\nu$ is the efficacy of cross-link~\cite{16}. Note the additional multiplicative factor of $2$, which arises because bisacrylamide is a tetra-functional cross-link. $\nu=1$ implies that all cross-links are effective, i.e. the polymer strands attached to each cross-link are a part of the homogeneous network. Any unproductive reaction of bisacrylamide or inhomogeneity in the network, for example if one of the four polymer strands connected to a cross-link is a dangling chain which does not contribute to the elasticity of the polymer network, leads to $\nu<1$~\cite{14,16,22}. Taking the molecular weight of a bisacrylamide as 154, the elasticity of a polyacrylamide gel at room temperature can be rewritten as $G'=33.2\times10^4\nu c$ measured in Pascals, where $c$ is the percentage concentration of bisacrylamide. In Fig.~\ref{fig:fig2}(b), we see that $G'$ is equal to $4.0\times10^4c~$Pa for polyacrylamide gels with 7.5\% acrylamide, and $3.0\times10^5c~$Pa for 15\% acrylamide respectively. Hence $\nu=0.12$ for the 7.5\% and $\nu=0.9$ for the 15\% PA gels. The higher value of $\nu$ for 15\% acrylamide PA gels is due to the higher polymer chain density in this system, suggesting that there is a higher probability for a bis molecule to find an acrylamide polymer chain in its neighborhood that would result in an effective cross-link.

Note though, that this is a simplified scenario; $\nu$ does not keep increasing indefinitely with increasing polymer chain concentration, but levels off as the semi-dilute limit for acrylamide chains is reached. The elastic modulus is also strongly affected by the amount of bis present, an excess of which may change the polymer solubility from good to theta solvent and the effective persistence length of the acrylamide chains~\cite{23}, and may even lead to macroscopic syneresis for sufficiently large concentrations of bis~\cite{24}. Thus, the relative concentrations of acrylamide and bis may have profound effect on the bulk modulus of PA gels, where instead of a linear scaling of the elastic modulus with the bis concentration, as seen in our samples (Fig.~\ref{fig:fig2}(b)), the elastic modulus will level off~\cite{25} due to micro-phase separation in the gels. The roles of bis and acrylamide are not limited to macroscopic elastic modulus only, but also have significant impact on the microscopic non-affinity of PA gels, as we note in the following section.

\subsection{Elastic Inhomogeneities in Polyacrylamide Gels}
The value of $\frac{\mathcal{A}}{\gamma^2}$ characterizes the inhomogeneities in the elasticity of the material.  In this section we analyze two possible scenarios that give rise to randomness in elastic modulus $K$ (i.e., randomness in the shear modulus $G'$ of PA gels), and thus generate non-affine responses in polyacrylamide gels.  We then compare the predicted non-affinity in these two scenarios with experimental results.

In the first scenario, the gel is assumed to be nearly ideal.  The inhomogeneity is assumed to be produced by the intrinsic randomness in the network geometry arising from thermal fluctuations frozen into the PA gel at the moment of gelation; in this case, the smallest lengthscale  characterizing the inhomogeneity would be the mesh size of the network, and the inhomogeneities cannot be reduced by improving the synthesis process.  In the second scenario, ``nonthermal'' inhomogeneities are assumed to be introduced during the sample preparation process. For example, this effect could arise if the cross-links are not homogeneously distributed; in this second case, elastic inhomogeneities could be present at lengthscales much larger than the network mesh size.

The elastic modulus correlation function $\Delta^{K}$ in the first scenario can be modeled as
\begin{eqnarray}
	\Delta^{K}(x) = (\delta G')^2 \xi_e^3 \, \delta(x) ,
\end{eqnarray}
where $(\delta G')^2$ is the variance of the local shear modulus $G'$, and $\xi_e$ is the characteristic mesh size of the network.  In this scenario the gelation process is nearly ideal, and the only randomness comes from the frozen thermal fluctuations in the liquid at the moment of gelation. Thus the correlation of the elastic modulus is characterized by the only length scale in the system, the mesh size $\xi_e$, which we also take to be the short-distance cutoff of the system because the picture of continuous elasticity breaks down below this length scale.  Thus the elastic modulus correlations at this scale is characterized by a Dirac delta-function in Eq.~(\ref{EQ:AInt}).  The Fourier transform of $\Delta^{K}$ is then
\begin{eqnarray}\label{EQ:Dirac}
	\Delta^{K}(q) = (\delta G')^2 \xi_e^3.
\end{eqnarray}
We plug this correlation function back to Eq.~(\ref{EQ:Dirac}) to calculate $\mathcal{A}$.  To evaluate the integral, one has to set a small length scale cutoff, which is $\xi$ as we discussed.  Below this length scale the polymer network structure cannot be coarse grained, and one cannot characterize the properties using continuous elasticity.

We obtain the non-affine parameter $\mathcal{A}$ from the integral (ignoring unimportant $O(1)$ constant prefactor):
\begin{eqnarray}
	\mathcal{A} \sim \Big(\frac{\delta G'}{G'}\Big)^2 \gamma^2 \xi_e^2.
	\label{eq:eq7}
\end{eqnarray}
 
The experimentally measured $\mathcal{A}$ is plotted as a function of $\gamma^2$ at two different monomer and cross-link concentrations in Fig.~\ref{fig:fig3}(a).  Both results are consistent with the theoretical prediction that $\mathcal{A}$ is proportional to $\gamma^2$.  However, the magnitude of $\frac{\mathcal{A}}{\gamma^2}$ is of the order of $1\mu m^2$. Since the mesh size is expected to be of order 10 nm, we obtain $\delta G'/G' \sim O(10^2)$ from Eq.~(\ref{eq:eq7}) and our measurements of $\frac{\mathcal{A}}{\gamma^2}$. This value is too large for a nearly ideal ``thermal'' gel.  In a nearly ideal gel, the inhomogeneities in network geometry and thus the elasticity come purely from thermal fluctuations at the moment of gelation, thus both $\delta G'$ and $G'$ are of order $k_B T$ times the cross-link number density, so one should expect $\delta G'/G' \sim O(1)$.  Furthermore, in this scenario $\mathcal{A}$ should be related to the concentration of cross-links $c$ as $\mathcal{A}\propto \xi_e^2\propto c^{-2/3}$ (because $\xi_e\propto c^{-1/3}$), but this behavior is not seen in Fig.~\ref{fig:fig3}(b), in which $\frac{\mathcal{A}}{\gamma^2}$ is essentially a constant (albeit with a wide scatter).

Other length scales in the gel will not significantly affect this analysis.  The persistence length of the polymer chain is even smaller than the mesh size, the small length cutoff of the analysis, and thus will not affect the result.  The size of the tracer bead, although on the scale of $1\mu m$ and relevant to the problem, only weakly changes the value of $\mathcal{A}$, as we discuss in Section~\ref{SEC:BEADSIZE} and Section 2 of the supporting material available online.

The discrepancy between the value of $\delta G'/G'$ suggested by the experiment and the theoretical value of nearly ideal ``thermal'' gel suggest that our second scenario may be more realistic for these systems.  In the second scenario, ``nonthermal'' inhomogeneities in the distribution of the bisacrylamide (cross-link) during the process of polymerization are assumed to exist.  These inhomogeneities are frozen in at polymerization and their contribution to the inhomogeneous elasticity in the resulting PA gel dominates over the contributions of thermal fluctuations at gelation described in the first scenario, because these ``nonthermal'' inhomogeneities exhibit greater variance and longer correlation length, as we discuss below.  These types of heterogeneities have been recognized previously in the literature~\cite{23,24,26,27,28,29,30,31,32,33,34,35,36,37,38}.  Briefly, because the hydrophobicity of polymerized bisacrylamide is higher than the polyacrylamide chains, the hydrophobic cross-links have a tendency to aggregate during the sample preparation~\cite{24,39}.  This effect can generate an inhomogeneous spatial distribution of cross-links at length scales longer than the mesh size.

Thus, in the second scenario, regions with high shear modulus and regions with low shear modulus form in the polyacrylamide network as a result of the inhomogeneous distribution of cross-links frozen in during the process of sample preparation.  We define the length scale $\xi_G$ to characterize the size of this inhomogeneity.  The resultant inhomogeneity in the shear modulus may then be characterized by a Gaussian correlation function,
\begin{eqnarray}\label{EQ:Deltar}
	\Delta^{K}(r) = (\delta G')^2 e^{-\frac{r^2}{2\xi_G}}.
\end{eqnarray}
We then plug this correlation function it back into Eq.~(\ref{EQ:AInt}). The integral is convergent due to the finite range of this correlation function, i.e., so the short length scale cutoff is not needed in this scenario.  
The resulting non-affine parameter is given by 
\begin{eqnarray}
	\mathcal{A} \sim \Big(\frac{\delta G'}{G'}\Big)^2 \gamma^2 \xi_G^2 .
\end{eqnarray}
For a careful derivation of this relation with the exact value of the prefactor, see Section 1 of the supporting information available online.

In fact, considerable effort has been expended over the years to characterize inhomogeneities inherent to PA gels. Starting with the pioneering work of Richards and Temple (1971)~\cite{28}, various experimental techniques, \textit{viz.}, gel-swelling and permeability studies~\cite{24,32,33}, small angle x-ray~\cite{23,24,26} and neutron scattering~\cite{23,27}, quasi-elastic light scattering~\cite{23,35,38}, dynamic light scattering~\cite{24,37}, UV-visible~\cite{34} and IR spectroscopy~\cite{31}, NMR spectroscopy~\cite{30,31}, electron micrographs~\cite{29}, have been used to quantify the nature and size of inhomogeneities created in PA gels.  Some of these ideas have been considered in the context of gel elastic properties~\cite{36}, as well as under varying acrylamide and bis concentration, and different polymerization reaction conditions. The ratio of monomer to cross-link concentrations, which determines the relative wettability of acrylamide and bis clusters during the polymerization process, as well as the reaction kinetics, all affect the formation of dense, heterogeneous clusters of highly cross-linked polymers interspersed with patches of sparsely cross-linked polymer chains. The size of these spatial inhomogeneties embedded in the more uniform gel matrix has been reported to vary widely from a few nanometers to as much as half a micron, with homogeneous regions of comparable length scale in between. 

One may substitute the inhomogeneity correlation length, $\xi_G$ with the size of the spatial inhomogeneities reported in the aforementioned references. From the literature we find that $5~nm \lesssim \xi_G\lesssim 500~nm$, which gives corresponding range of inhomogeneity magnitude of $3 \lesssim \frac{\delta G'}{G'} \lesssim 300$ for PA gels over a wide range of monomer and cross-link concentrations. For PA gels synthesized under similar preparation conditions as in our experiment, the length scale of inhomogeneities has been measured using a nano-indentation method, leading to $\xi_G \lesssim 200~nm$~\cite{40}, from which we obtain $	\frac{\delta G'}{G'} \lesssim 7$.

\subsection{Repeated Shear Measurements}
\label{REPEATED}
As part of this study, we explored the effects of cycled measurements on $\mathcal{A}$ in the same sample.  By repeatedly shearing and unshearing a PA gel sample at the same strain, we determined the distribution of $\mathcal{A}$ for the same set of particles within a single sample.  The resultant variation of $\mathcal{A}$ is not insignificant, though it is considerably less than sample-to-sample error. 

\begin{figure}[h!]
\subfigure[]{\includegraphics[trim = 35mm 85mm 35mm 85mm, clip, width=0.5\textwidth]{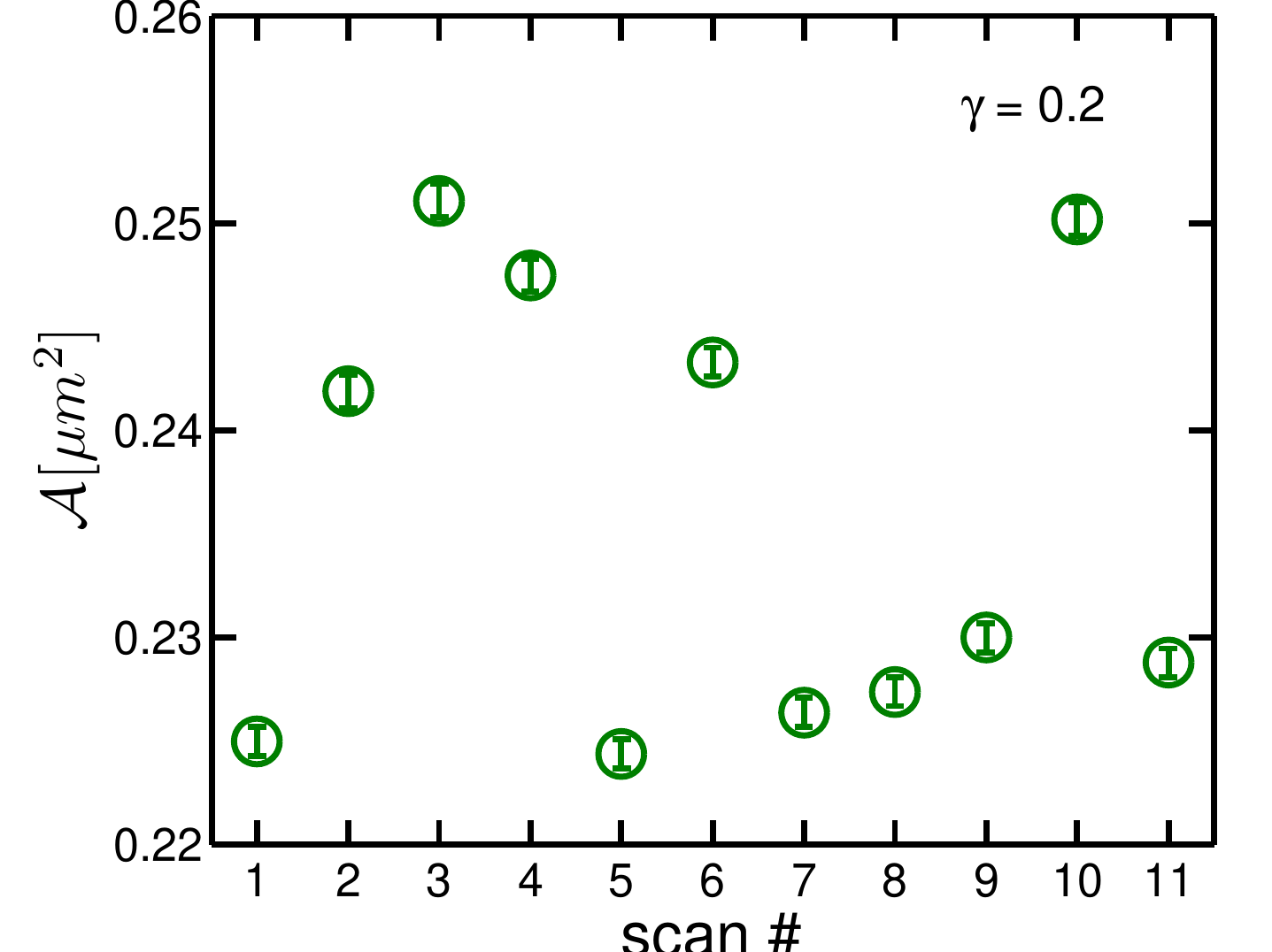}}
\subfigure[]{\includegraphics[trim = 35mm 80mm 45mm 85mm, clip, width=0.45\textwidth]{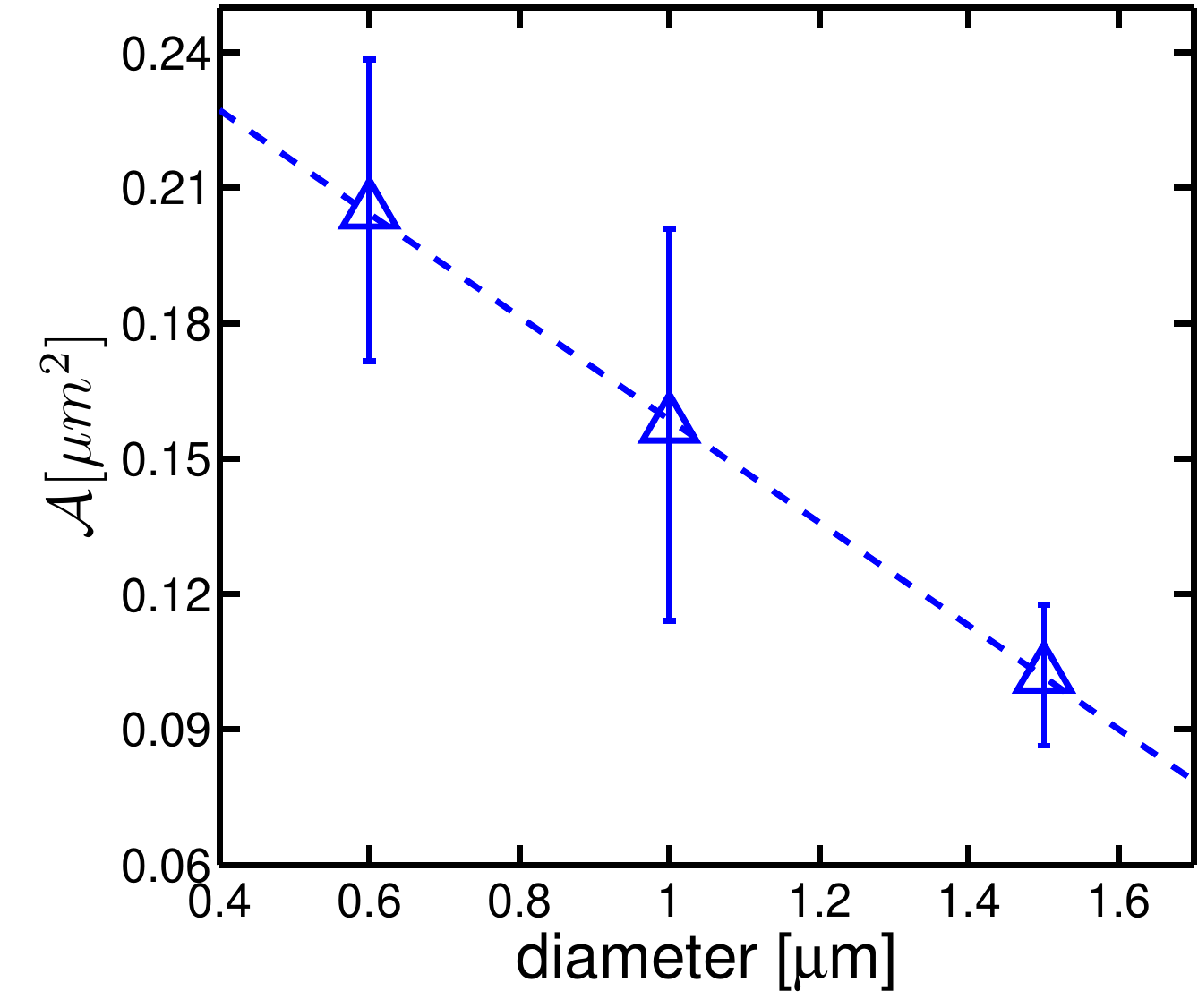}}\\
\subfigure[]{\includegraphics[trim = 30mm 75mm 30mm 75mm, clip, width=0.5\textwidth]{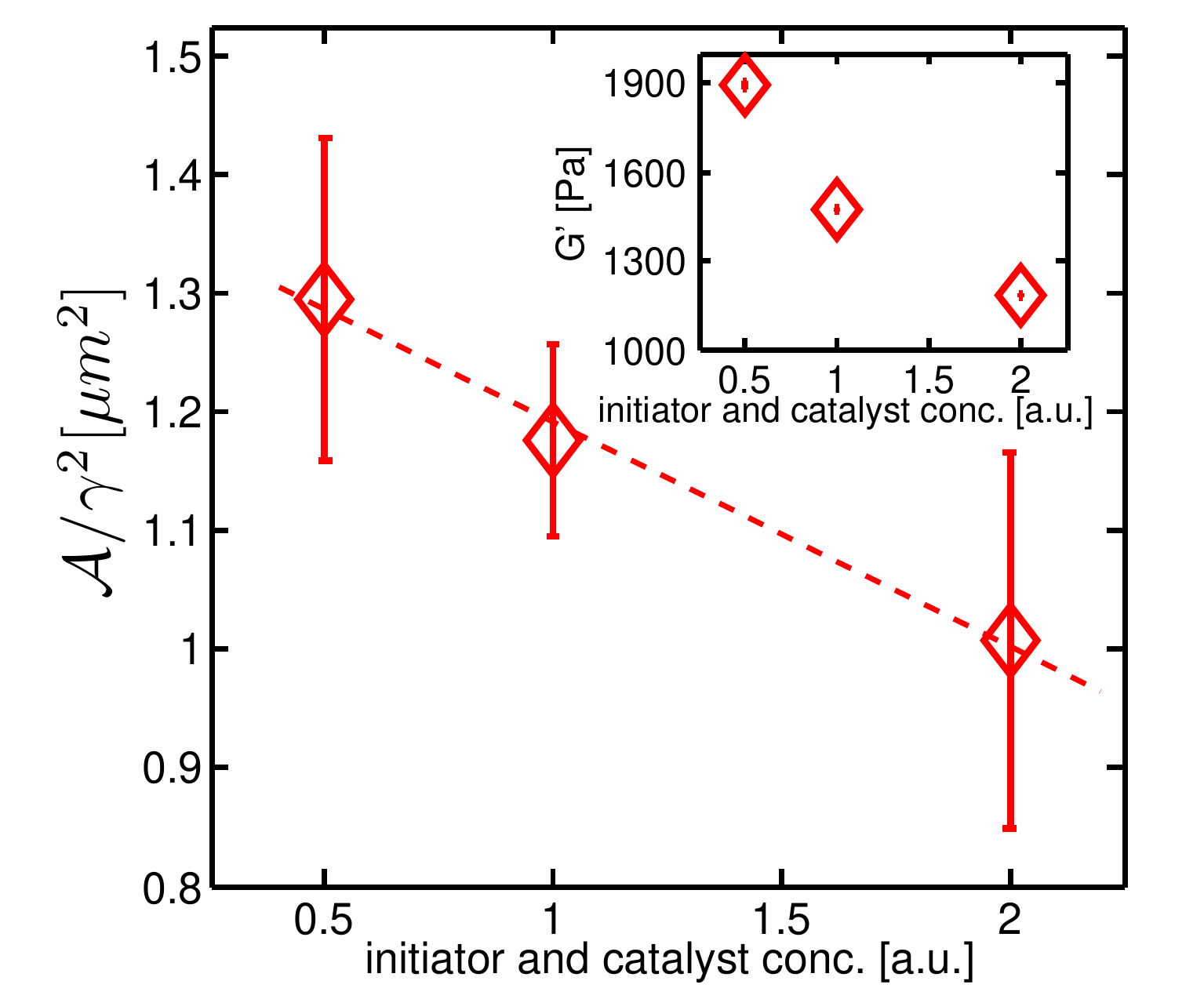}}
\caption{\label{fig:fig4} (a) Non-affine parameter, $\mathcal{A}$ in a sample PA gel (7.5\% acrylamide and 0.06\% bis), sheared repeatedly under $\gamma = 0.2$ strain. (b) Average non-affine parameter in a sample PA gel measured using fluorescent tracer beads of average diameters of $600~nm$, $1~\mu m$ and $1.5~\mu m$. $\mathcal{A}$ decreases linearly with the size of tracer beads. Measurements shown here were performed on a sample PA gel with 7.5\% acrylamide and 0.06\% bis, sheared ten times at a strain of $\gamma=0.3$. (c) Elastic shear modulus decreases with increasing initiator and catalyst concentrations, for PA gel where the monomer and cross-link concentrations have been kept constant(inset). $\mathcal{A}$ decreases linearly with increasing initiator and catalyst concentrations.  Data are shown here for a 7.5\% acrylamide and 0.03\% bis PA gel.}
\end{figure}

To demonstrate this effect, a PA gel sample is synthesized at 7.5\% acrylamide and 0.06\% bis with $1~\mu m$ tracer beads embedded in it. The gel is sheared repeatedly to a strain of $0.2$, and $\mathcal{A}$ is measured each time as shown in Fig.~\ref{fig:fig4}(a). Error bars reflect the systematic error in our measurements. The tracer beads relax back roughly to their original (unsheared) positions once the strain is released. The variation in $\mathcal{A}$ suggests that some local rearrangement of the polymer network neighborhood occurs after/during each cycle, perhaps because of the presence of compliant chain entanglements or reorganization of the gel-bead interface. These rearrangements permit the tracer beads to explore and experience slightly different local environments every time the sample undergoes a shear transformation. Non-affinity was slightly different after each shear event. The measured standard deviation of $\frac{\mathcal{A}}{\gamma^2}$ ($\sim 8\%$) for repetitive shear in the sample is much smaller, however, than that measured for different gels prepared under apparently identical experimental conditions.    

With respect to non-affinity variation with repeated cyling, we have explored this phenomenon under different strains as well as for different polymer gel concentrations. It appears that the randomness persists even when a sample gel is sheared repeatedly thirty times. The variation in non-affinity parameter appears to be random, independent of the number of times the gel is sheared. Chain entanglements, dangling ends, etc. could contribute to this randomness in the measured non-affinity~\cite{22}, and  one cannot rule out the possibility that the local environment of the tracer micro-beads is subtly distorted due to polymer depletion or adsorption, which might cause more/less slippage or sticking of the tracer beads to the surrounding gel matrix under shear~\cite{40,41}. We use this repeated shear technique to calculate the systematic error in our measurements to be $\sim 8\%$ an use this value as the lower bound for all error estimations shown in Fig.~\ref{fig:fig3}(b).

\subsection{Tracer Bead-size Dependence}\label{SEC:BEADSIZE}
We also explored the effects of the size of the tracer beads on the magnitude of the non-affine parameter, $\mathcal{A}$, using tracer beads of three different sizes, \textit{viz.}, 0.6, 1 and 1.5 $\mu m$. The different-sized beads are fluorescently labeled such that they are uniquely excited by three different wavelengths of the confocal scanning beam, \textit{viz.}, 488, 568, and 640 nm respectively. We disperse these three different-sized beads in a sample PA gel and image them using three different wavelength excitation beams in succession during a particular shear event. We see that, for the range of bead-sizes used in our experiment, the magnitude of $\mathcal{A}$ remains within the range indicated in Table~\ref{TAB:Summary} for 7.5\% acrylamide PA gels. We also note that though weak, there is a slight bead-size dependence of the average value of $\mathcal{A}$ measured from repeated shear events. Fig.~\ref{fig:fig4}(b) plots the average $\mathcal{A}$ from eleven repeated shear events at $\gamma=0.3$, for a PA gel at 7.5\% acrylamide and 0.06\% bis. We see that the average $\mathcal{A}$ decreases with the diameter of tracer beads. $\mathcal{A}$ can be fit to a linear function of tracer bead size, with a slope of $\sim -0.11\pm 0.001 \mu m^2$ and an intercept of $\sim 0.27 \pm 0.002$.
 
Essentially all of the theoretical analysis presented in this paper thus far employed the simplifying assumption that we can treat the tracer beads as point objects that probe local non-affine deformations.  However, the size of the tracer bead is comparable to the correlation length $\xi_G$ of the random elastic modulus.  In Section 2 of the supporting material available online, we compute the corrections due to the finite size $R$ of the bead in a simplified model of electrostatics in random media, which a scalar analogy to the elastic problem.  In the limit of $R\to 0$ the non-affine parameter $\mathcal{A}$ smoothly approaches the limit of point probe, while in the limit of $R/\xi\gg 1$, $\mathcal{A}$ approaches a different value which is related to the $R\to 0$ value by a constant factor of $O(1)$.  This simple calculation is consistent with the experimental observation (Fig.~\ref{fig:fig4}(b)) that $\mathcal{A}$ only weakly depends on bead-size, $R$.  However the exact dependence is not captured by the calculation.

\subsection{Effects of Initiator and Catalyst Concentration}
Finally, we explored the effects of reaction kinetics on the strain-normalized non-affinity measure, $\frac{\mathcal{A}}{\gamma^2}$. To do that, we prepare PA gels with same amount of monomer and cross-link concentration, viz., 7.5\% acrylamide and 0.03\% bis, but with the initiator and catalyst (TEMED and APS, respectively) concentrations twice and half of the normal amount used. The gel reactions proceed faster (slower) as a result, respectively, yielding lower (higher) plateau shear modulus for twice (half) the normal initator and catalyst concentrations (inset in Fig.~\ref{fig:fig4}(c)). The $\frac{\mathcal{A}}{\gamma^2}$ values calculated for these samples are still within range of $1.65~\pm 0.63~\mu m^2$, the measured average for 7.5\% acrylamide PA gels, leading us to believe that measured values of $\frac{\mathcal{A}}{\gamma^2}$ are still dominated by the inhomogeneities in these gels. Error bars reflect the standard deviation in the ensempble-averaged non-affinity values measured for four scans over each sample volume.   Within this prescribed range, though, there is a slight inverse dependence of $\frac{\mathcal{A}}{\gamma^2}$ on the concentration of TEMED and APS  (Fig.~\ref{fig:fig4}(c)) which we do not understand.

\section{Conclusions}

Non-affine deformations under shear are measured in a simple cross-linked gel and are employed to provide insight about inhomogeneities in the flexible polymer gels. Results indicate that the non-affine parameter, $\mathcal{A}$, which is the mean square non-affine deviation in the PA gels, is proportional to the square of the strain applied, in agreement with theoretical predictions based on a linear elastic theory treating disorder as perturbations.  Interestingly, the magnitude of $\mathcal{A}$ is greater than what one would expect from theoretical calculations assuming that the PA gels are nearly-ideal and the only source of disorder is from the frozen-in thermal fluctuations at gelation.  Furthermore, the degree of non-affinity appears to be independent of polymer chain density and cross-link concentration.  Thus, we posit that there are additional built-in inhomogeneities in the PA gels that lead to the large non-affininity we observe.  Indeed, there is ample evidence in existing literature of the presence of such inhomogeneities in PA gels due to a difference in the hydrophobicities of the bisacrylamide and acrylamide monomers.  We also discuss the length scale and magnitude of these additional inhomogeneities.  Our measurements of non-affinity in PA gels, which are model flexible polymer gels, provide a benchmark for the degree of non-affinity in soft materials, and will serve as am interesting comparison to non-affinity in more complicated materials such as semi-flexible bio-polymer networks.

\acknowledgement
This work was supported by the MRSEC DMR-0520020, DMR-0505048, and DMR-0079909 grants.

%

\appendix
\section{Non-affine correlation function in inhomogeneous media}\label{APP:AGPREF}

We assume that the gel has a homogeneous bulk modulus, but inhomogeneous shear modulus.  The inhomogeneity in shear modulus originates in the inhomogeneity of the cross-links due to effects such as hydrophobicity, etc. at the moment of gelation, while the monomers are relatively homogeneous, and thus the bulk modulus is homogeneous.

The mean and variance of the shear modulus can be written as
\begin{eqnarray}
	\langle G(r) \rangle &=& G \nonumber\\
	\langle G(r)G(r') \rangle -G^2 &=& \langle \delta G(r) \delta G(r') \rangle = (\delta G)^2 e^{-\frac{\vert r-r' \vert^2}{2\xi^2}},
\end{eqnarray}
where $\langle\cdots\rangle$ represent disorder average, $\xi$ is the correlation length of the shear modulus, and $\delta G$ represent the strength of the disorder in $G$.  In this Appendix we drop the $'$ on shear modulus $G'$ for convenience of notation. 

To calculate the non-affine deformation, we use the form of the elastic energy, Eq.(3.25), in Ref.~\cite{2}, which is a first order expansion in small randomness in elastic constant $K_{ijkl}$ and the resulting small non-affine deformation $u$
\begin{eqnarray}
	\delta \mathcal{H} = \frac{1}{2} \int \big\lbrace K_{ijkl} \partial_j u_i \partial_l u_k 
	+ \tilde{\sigma}_{jl}(r)  \partial _j u_i \partial_l u_i
	+2\delta K_{ijkl} (r) \gamma_{kl} \partial_j u_i \big\rbrace ,
\end{eqnarray}
where we've only kept first order terms.  For isotropic media the elastic constant $K_{ijkl}$ is related to the Lame coefficients as
\begin{eqnarray}
	K_{ijkl} = \lambda \delta_{ij}\delta_{kl} + G (\delta_{ik}\delta_{jl} + \delta_{il}\delta_{jk}),
\end{eqnarray}
in which $G$ is also the shear modulus, and the bulk modulus is given by $B=\lambda+2G$.

The variation in $u$ to minimize $\delta \mathcal{H} $ gives to leading order
\begin{eqnarray}
	- K_{ijkl} \partial_j\partial_l u_k(r) = \partial _j \delta K_{ijkl}(r) \gamma_{kl},
\end{eqnarray}
where the $\tilde{\sigma}$ term vanishes at equilibrium.  This equation can be written in momentum space as
\begin{eqnarray}
	K_{ijkl} q_j q_l u_k(q) = -q_j \delta K_{ijkl}(q) \gamma_{kl}.
\end{eqnarray}
We define the driving force $f_i(q) \equiv -q_j \delta K_{ijkl}(q) \gamma_{kl}$, then the above equation can be written as
\begin{eqnarray}\label{EQ:Pert}
	\big\lbrack (\lambda+2G) \mathcal{P}^{L}(q) +G \mathcal{P}^{T}(q) \big\rbrack _{ik} q^2 u_k(q) = f_i(q),
\end{eqnarray}
where $\mathcal{P}^{L}_{ik}(q) \equiv q_i q_k /q^2$ and $\mathcal{P}^{L}_{ik}(q) \equiv \delta_{ik}-q_i q_k /q^2$ are the projection operators, and they are orthogonal.  Thus the solution of the non-affine deformation as a function of given disorder $\delta K$ is
\begin{eqnarray}
	u_i(q) = \Bigg\lbrack \frac{1}{\lambda+2G} \mathcal{P}^{L}(q) + \frac{1}{G} \mathcal{P}^{T}(q)
	\Bigg\rbrack _{ik}  \frac{1}{q^2} f_k(q) .
\end{eqnarray}

Using this we can calculate the correlation function of the non-affine deformation $\mathcal{G}(r,r')\equiv \langle u_m(r) u_m(r') \rangle$ in Fourier space

\begin{eqnarray}
	\mathcal{G}(q,q') &=& \langle u_m(q) u_m(q') \rangle \nonumber\\
	&=& \Bigg\lbrack \frac{1}{\lambda+2G} \mathcal{P}^{L}(q) + \frac{1}{G} \mathcal{P}^{T}(q)
	\Bigg\rbrack _{mi} \Bigg\lbrack \frac{1}{\lambda+2G} \mathcal{P}^{L}(q) + \frac{1}{G} \mathcal{P}^{T}(q)
	\Bigg\rbrack _{mi'} \frac{1}{q^4} \langle f_i(q)f_{i'}(q') \rangle \nonumber\\
	&=& \Bigg\lbrack \frac{1}{(\lambda+2G)^2} \mathcal{P}^{L}(q) + \frac{1}{G^2} \mathcal{P}^{T}(q) \Bigg\rbrack _{ii'}\frac{1}{q^4} 
	q_j q_{j'} \langle \delta K_{ijkl}(q)\delta K_{i'j'k'l'}(q')\rangle \gamma_{kl}\gamma_{k'l'} .
\end{eqnarray}
We assume that there's only disorder in shear modulus $G$, so the correlation function is (as discussed in Appendix A and B in Ref.~\cite{2})
\begin{eqnarray}
	\langle \delta K_{ijkl}(q)\delta K_{i'j'k'l'}(q')\rangle 
	= (\delta_{ik}\delta_{jl} + \delta_{il}\delta_{jk})(\delta_{i'k'}\delta_{j'l'} + \delta_{i'l'}\delta_{j'k'}) \langle \delta G(q)\delta G(q')\rangle .
\end{eqnarray}
As we already discussed,
\begin{eqnarray}
	\langle \delta G(q)\delta G(q')\rangle = (2\pi)^3 \delta(q+q') (\delta G)^2 (2\pi)^{3/2} \xi^3 e^{-\frac{\xi^2 \vert q\vert^2}{2}},
\end{eqnarray}
and suppose the shear deformation has the form ($\gamma$ is defined as $\Lambda=I+\gamma$)
\begin{eqnarray}
	\gamma = \left(
		\begin{array}{ccc}
		0 & 0 & \gamma_{xz} \\
		0 & 0 & 0 \\
		0 & 0 & 0 
		\end{array}
	\right),
\end{eqnarray}
we arrive at the correlation function
\begin{eqnarray}
	\mathcal{G}(q,q') = (2\pi)^3 \delta(q+q') \mathcal{G}(q)
\end{eqnarray}
with
\begin{eqnarray}
	\mathcal{G}(q)  = \Bigg\lbrack \frac{4\hat{q}_x^2\hat{q}_z^2}{(\lambda+2G)^2 q^2}  
	+ \frac{\hat{q}_x^4 +\hat{q}_x^2 \hat{q}_y^2 -2 \hat{q}_x^2\hat{q}_z^2 +\hat{q}_y^2\hat{q}_z^2 +\hat{q}_z^4}{G^2 q^2} 
	 \Bigg\rbrack 
	 \gamma_{xz}^2  (\delta G)^2 (2\pi)^{3/2} \xi^3 e^{-\frac{\xi^2 \vert q\vert^2}{2}}.
\end{eqnarray}
Using this we Fourier transform back to real space and get $\mathcal{A}$ as
\begin{eqnarray}\label{EQ:ElasCorr}
	\mathcal{A} &\equiv & \langle \delta u_m(r) \delta u_m (r) \rangle \nonumber\\
	&=& \int \frac{d^3 q }{(2\pi)^3} \mathcal{G}(q) \nonumber\\
	&=& \gamma_{xz}^2 (\delta G)^2 (2\pi)^{3/2} \xi^3 \int \frac{d^3 q}{(2\pi)^3} \Bigg\lbrack \frac{4\hat{q}_x^2\hat{q}_z^2}{(\lambda+2G)^2 q^2}  
	+ \frac{\hat{q}_x^4 +\hat{q}_x^2 \hat{q}_y^2 -2 \hat{q}_x^2\hat{q}_z^2 +\hat{q}_y^2\hat{q}_z^2 +\hat{q}_z^4}{G^2 q^2} 
	 \Bigg\rbrack  e^{-\frac{\xi^2 \vert q\vert^2}{2}} \nonumber\\
	&=& \gamma_{xz}^2 (\delta G)^2 (2\pi)^{-3/2} \xi^3 \int q^2 \sin\theta dq d\theta d\phi \nonumber\\
	&&\quad\times e^{-\frac{\xi^2 \vert q\vert^2}{2}}\Bigg\lbrace \frac{4(\sin\theta\cos\phi\cos\theta)^2}{(\lambda+2G)^2 q^2}  
	 + \frac{1}{G^2q^2} \Big\lbrack (\sin\theta\cos\phi)^4 + (\sin\theta\cos\phi\sin\theta\sin\phi)^2 
	 	\nonumber\\
	&&\quad\,	+ (\sin\theta\sin\phi\cos\theta)^2
				-2(\sin\theta\cos\phi\cos\theta)^2
				+(\cos\theta)^4
	 \Big\rbrack
	 \Bigg\rbrace  \nonumber\\
	&=& \gamma_{xz}^2 (\delta G)^2 (2\pi)^{-1} \xi^2 \Big\lbrack \frac{1}{(\lambda+2G)^2} \frac{16\pi}{15}+\frac{1}{G^2} \frac{8\pi}{5}\Big\rbrack \nonumber\\
	&=& \frac{4}{5}\gamma_{xz}^2 \frac{(\delta G)^2}{G^2}  \xi^2 + \frac{8}{15}\gamma_{xz}^2 \frac{(\delta G)^2}{(\lambda+2G)^2}  \xi^2 
\end{eqnarray}
The second term may be small compared to the first term, given that the bulk modulus is large.

\section{Corrections for finite-sized bead}\label{APP:FINITESIZE}
In this experiment, local displacements are probed by beads of size in the order of $1~\mu m$ that are embedded in the media, and the displacement of a bead can be different from that of a point probe (the position of the point probe is the same as the center of the bead).  The correction due to the finite size of the bead is of order $\xi_G/R$ where $R$ is the radius of the bead, proved as follows.

We simplify the discussion by ignoring the vector nature of the elastic displacements, and consider the analogous scalar model that studies a finite sized conductor bead in a inhomogeneous electric field.  The correspondence of the quantities in electrostatics and elasticity are listed in Table~\ref{TAB:Corr}.

Let us first consider the simplest case, that a conductor bead of radius $R$ placed at the origin $\mathbf{r}=0$ in a homogeneous external field $E\mathbf{e}_{z}$, the electric potential $\phi_0$ outside the bead satisfies the Laplace's equation
\begin{eqnarray}
	\nabla ^2 \phi_0 (\mathbf{r}) = 0 ,
\end{eqnarray}
and the solution is well known
\begin{eqnarray}\label{EQ:zero}
	\phi_0(\mathbf{r}) = E r \cos\theta - E \Big(\frac{R}{r}\Big)^3 r \cos\theta ,
\end{eqnarray}
where $\theta$ is the inclination angle of $\mathbf{r}$ measured from $\mathbf{e}_{z}$.  The potential of the bead itself is $\phi_0\equiv\phi_0(\mathbf{r})_{\vert \mathbf{r}\vert=R}=0$.

\begin{table}
\begin{tabular}{|c|c|}
\hline
Electrostatics & Elasticity \\
\hline
$\epsilon$ (permittivity) & $K$ (elastic modulus) \\
$\phi$ (potential) &  $u$ (displacement) \\ 
$E$ (field) & $\gamma$ (strain) \\
\hline
\end{tabular}
\caption{Analogy between quantities in electrostatics (scalar problem) and elasticity (vector problem) of a finite sized bead placed in an inhomogeneous media in with external field.}
\label{TAB:Corr}
\end{table}

For the case of inhomogeneous media, which has randomness in the permittivity $\epsilon(\mathbf{r}) = \epsilon_0 + \delta\epsilon(\mathbf{r})$ that results in a random potential $\phi(\mathbf{r})=\phi_0(\mathbf{r}) +\delta\phi(\mathbf{r})$, the Laplace's equation is
\begin{eqnarray}
	\nabla \cdot \lbrack (\epsilon_0 + \delta\epsilon(\mathbf{r}))\nabla (\phi_0 +\delta\phi(\mathbf{r}))\rbrack =0 .
\end{eqnarray}
This equation can be solved perturbatively, and the zero-th order solution $\phi_0$ is given in Eq.~(\ref{EQ:zero}).  The first order perturbation equation is
\begin{eqnarray}
	\epsilon_0 \nabla^2 \delta\phi(\mathbf{r}) = -(\nabla \delta\epsilon(\mathbf{r})) \cdot (\nabla \phi_0(\mathbf{r})),
\end{eqnarray}
for the electric potential outside the bead ($r\equiv\vert \mathbf{r}\vert>R$), and the boundary condition is that at the surface of the conductor bead $r=R$, the potential $\delta\phi(\mathbf{r})=\textrm{constant}$.  This equation is analogous to the first order equation, Eq.~(\ref{EQ:Pert}) for the elasticity case.  This electrostatic equation can be solved using the Green's function defined as
\begin{eqnarray}
	\nabla^2 \mathscr{G}(\mathbf{r},\mathbf{r'}) = -\frac{1}{\epsilon_0} \delta(\mathbf{r}-\mathbf{r'}),
\end{eqnarray}
determining the potential at point $\mathbf{r}$ with a point unit charge $q=1$ at point $\mathbf{r}'$ (outside the bead), and the boundary condition is that $\mathscr{G}(\mathbf{r})=\textrm{constant}$ at $r=R$.  This can be solved using the method of image charges, and we obtain the following solution
\begin{eqnarray}
	\mathscr{G}(\mathbf{r},\mathbf{r'}) 
	= \frac{1}{4\pi \epsilon_0} \Big( \frac{1}{\vert \mathbf{r}-\mathbf{r'}\vert} -\frac{q'}{\vert \mathbf{r}-\mathbf{r''}\vert}+\frac{q'}{\vert \mathbf{r}\vert}\Big)
\end{eqnarray}
where $q'= R/\vert \mathbf{r}'\vert$ and $\mathbf{r''}=\mathbf{r'}R^2/\vert \mathbf{r}'\vert^2$.  We are actually only interested in the potential of the bead itself, which is
\begin{eqnarray}
	\tilde{\mathscr{G}}(\mathbf{r'}) &\equiv& \mathscr{G}(\mathbf{r},\mathbf{r'}) \vert_{\vert \mathbf{r}\vert=R} \nonumber\\
	&=& \frac{1}{4\pi \epsilon_0} \frac{q'}{R} \nonumber\\
	&=& \frac{1}{4\pi \epsilon_0 \mathbf{r'}} 
\end{eqnarray}
which equals to the potential at the point $\mathbf{r}=0$ (the center of the bead) if there were no bead placed there.

Then using this Green's function we calculate the first order part of the potential $\delta \phi$ of the bead,
\begin{eqnarray}
	\delta\phi \equiv \delta\phi(\mathbf{r})\vert_{\vert \mathbf{r}\vert=R} 
	= \int_{\vert \mathbf{r}'\vert>R} d\mathbf{r} \frac{1}{4\pi \epsilon_0\vert\mathbf{r'}\vert} (\nabla \delta\epsilon(\mathbf{r}'))\cdot (\nabla \phi_0(\mathbf{r}')),
\end{eqnarray}
Thus the correlation function of the potential $\delta\phi$ is determined by the correlation function of the random permittivity $\epsilon(\mathbf{r})$ as
\begin{eqnarray}
	\langle \delta\phi^2 \rangle = \int_{\vert \mathbf{r}_1\vert>R,\vert \mathbf{r}_2\vert>R} d\mathbf{r}_1 d\mathbf{r}_2 
	\frac{1}{(4\pi \epsilon_0)^2\vert\mathbf{r}_1\vert\vert\mathbf{r}_2\vert} \partial_{1,i} \phi_0(\mathbf{r}_1)\partial_{2,j} \phi_0(\mathbf{r}_2)
	\partial_{1,i} \partial_{2,j}\langle \delta\epsilon(\mathbf{r}_1)\delta\epsilon(\mathbf{r}_2) \rangle ,
\end{eqnarray}
where $\partial_{1,i}\equiv\partial/\partial r_{i,i}$.  
Assuming that $\langle \delta\epsilon(\mathbf{r}_1)\delta\epsilon(\mathbf{r}_2) \rangle=(\delta \epsilon)^2 C(\mathbf{r}_1-\mathbf{r}_2)$ where $C(\mathbf{r}_1-\mathbf{r}_2)$ has the Gaussian form $e^{-\frac{\vert\mathbf{r}_1-\mathbf{r}_2\vert^2}{2\xi^2}}$ as we assumed in the elasticity case.  
This integral can be evaluated in momentum space as
\begin{eqnarray}
	\langle \delta\phi^2 \rangle = \frac{(\delta \epsilon)^2}{\epsilon_0^2} \int\frac{d\mathbf{q}}{(2\pi)^3} F(\mathbf{q})F(-\mathbf{q})C(\mathbf{q})
\end{eqnarray}
with
\begin{eqnarray}
	C(\mathbf{q}) \equiv \int d\mathbf{r} C(\mathbf{r}) e^{i\mathbf{q}\mathbf{r} } = (2\pi\xi^2)^{3/2} e^{-\xi^2 \vert\mathbf{q}\vert^2/2}
\end{eqnarray}
and
\begin{eqnarray}
	F(\mathbf{q})\equiv \int_{\vert \mathbf{r}\vert>R} d\mathbf{r} \frac{i\mathbf{q}\cdot\mathbf{E}_0(\mathbf{r})e^{i\mathbf{q}\mathbf{r} }}{4\pi\vert \mathbf{r}\vert }
\end{eqnarray}
where $\mathbf{E}_0\equiv\nabla\phi_0$ is the electric field in the zero-th order solution.  With some tedious calculation we obtain
\begin{eqnarray}
	F(\mathbf{q}) = \frac{i q_z E}{q^2} f(qR)
\end{eqnarray}
with
\begin{eqnarray}
	f(qR)\equiv	\cos(qR) -\frac{2(6-q^2R^2)\sin(qR)-2qR(6+q^2R^2)\cos(qR)+q^4R^4(\pi-2Si(qR))}{8qR}
\end{eqnarray}
where $q=\vert\mathbf{q}\vert$ and $Si$ is the Sine Integral.  Therefore $\langle \delta\phi^2 \rangle$ can be calculated in spherical coordinate as
\begin{eqnarray}
	\langle \delta\phi^2 \rangle = \frac{(\delta \epsilon)^2}{\epsilon_0^2} \frac{E^2(2\pi)^{3/2}\xi^3}{6\pi\xi} \int_{0}^{\infty} dx e^{-x^2/2}  f\Big(x\frac{R}{\xi}\Big)^2
\end{eqnarray}
where $x=\xi q$.  Numerical calculation shows that this integral as a function of $R/\xi$ has plateaus of value $\sqrt{\pi/2}$ at small $R/\xi$, at which $\langle \delta\phi^2  \rangle$ reduces to the value of point probes, and of value $\sim 5.6$ at large $R/\xi$, with slowly varying values in between these two limits.  Thus, the correlation of the potential is
\begin{eqnarray}
	\langle \delta\phi^2 \rangle = b \frac{(\delta \epsilon)^2}{\epsilon_0^2} E^2 \xi^2,
\end{eqnarray}
where $b$ is is a constant of $O(1)$ with a weak dependence on $R/\xi$.  This is consistent with the experimental finding that the non-affine parameter only weakly depends on $R/\xi$, although the dependence on $R/\xi$ is not correctly captured by the theory.

\section{Correlation of non-affine displacements between different beads}

The correlation of the non-affine displacements between difference beads decay with distance quickly as $1/r$, and is thus not visible in the current experiment, as discussed below.

The real space non-affine correlation function is the Fourier transform of $\mathcal{G}(q)$ as given in Eq.~(3) in the main paper.  Using the correlation function for the inhomogeneous shear modulus as given in Eq.~(8), again in the main paper, we get
\begin{eqnarray}
	\mathcal{G}(r) = \gamma^2 \Big(\frac{\delta G'}{G'}\Big)^2 \xi^3 \sqrt{\frac{\pi}{2}}\, \frac{1}{r}\,
	 \textrm{Erf}\Big(\frac{r}{\sqrt{2}\xi}\Big) ,
\end{eqnarray}
where Erf is the error function.  In the limit $r\to 0$, 
\begin{eqnarray}
	\mathcal{G}(r)\vert_{r\to 0} = \gamma^2 \Big(\frac{\delta G'}{G'}\Big)^2 \xi^2 = \mathcal{A}
\end{eqnarray}
reduces to the same point non-affine correlation function $\mathcal{A}$ which is the non-affine parameter we use.  In the limit $r \gg \xi$
\begin{eqnarray}
	\mathcal{G}(r)\vert_{r\to \infty} = \gamma^2 \Big(\frac{\delta G'}{G'}\Big)^2 \xi^3 \sqrt{\frac{\pi}{2}} \frac{1}{r} \sim \frac{1}{r}.
\end{eqnarray}
In this experiment, the distance between the beads is greater than the diameter of the bead ($1~\mu m$), given that $\xi$ is about $0.1~\mu m$, we have
\begin{eqnarray}
	\mathcal{G}(r) \simeq \sqrt{\frac{\pi}{2}} \frac{\xi}{r} \mathcal{A} < 0.1 \mathcal{A},
\end{eqnarray} 
and decays as $r$ increases.  

Indeed, for beads separated by 500 nm or more, as in our case, we do not detect any correlations in the non-affine displacements between different beads. This result is also in agreement with what is observed in the numerical results in Ref.~\cite{2}.

\section{Ensemble averaged $\frac{\mathcal{A}}{\gamma^2}$ values for PA gel samples}\label{APP:ENSEMBLE}

\begin{figure}[htp]
\subfigure[]{\includegraphics[trim = 35mm 85mm 35mm 85mm, clip, width=0.45\textwidth]{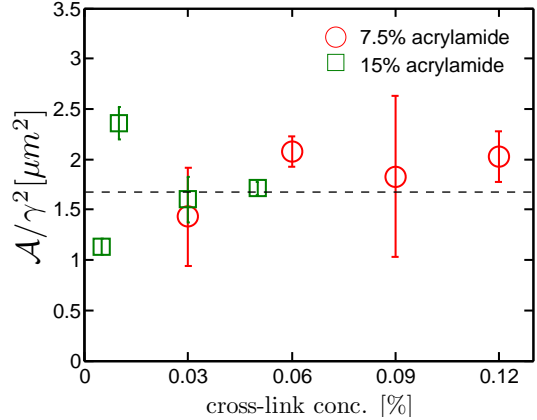}}
\caption{\label{fig:fig5}(a) $\frac{\mathcal{A}}{\gamma^2}$ for sample PA gels at 7.5\% and 15\% acrylamide are plotted at varying bis concentrations. The data points represent the ensemble-averaged values of measurements from several different samples prepared, ostensibly, in the same manner. Error bars reflect the fitting error in calculating ensemble-averaged $\frac{\mathcal{A}}{\gamma^2}$. The dashed line in the main figure indicates the mean of all $\frac{\mathcal{A}}{\gamma^2}$ values listed in Table~\ref{TAB:List}.}
\end{figure}

\begin{table}
\begin{tabular}{|c c c c c|}
\hline
Sample\#$~$ & acrylamide conc.[\%]$~~$& bis conc.[\%]$~~$&  $\frac{\mathcal{A}}{\gamma^2}~[\mu m^2]~~~$  &  Error [$\mu m^2$]\\
\hline 
1 & 7.5 & 0.03 & 1.18 & $\pm~$0.08\\ 
2 & 7.5 & 0.03 & 1.92 & $\pm~$0.96\\ 
3 & 7.5 & 0.06 & 2.08 & $\pm~$0.15\\ 
4 & 7.5 & 0.06 & 0.61 & $\pm~$0.28\\ 
5 & 7.5 & 0.09 & 1.15 & $\pm~$0.58\\ 
6 & 7.5 & 0.09 & 2.03 & $\pm~$0.94\\ 
7 & 7.5 & 0.12 & 1.69 & $\pm~$0.50\\ 
8 & 7.5 & 0.12 & 2.72 & $\pm~$0.25\\
9 & 7.5 & 0.12 & 1.50 & $\pm~$0.41\\  
10 & 15 & 0.005 & 1.13 & $\pm~$0.08\\ 
11 & 15 & 0.01 & 2.36 & $\pm~$0.16\\ 
12 & 15 & 0.03 & 1.60 & $\pm~$0.23\\ 
13 & 15 & 0.05 & 1.71 & $\pm~$0.07\\ 
\hline
\end{tabular}
\caption{List of $\frac{\mathcal{A}}{\gamma^2}$ values for different PA gel samples. Error estimates reflect the uncertainty in the linear fits from which $\frac{\mathcal{A}}{\gamma^2}$ are obtained. }
\label{TAB:List}
\end{table}

Fig.~\ref{fig:fig5}(a) plots the ensemble average of the strain-normalized non-affinity parameter, $\frac{\mathcal{A}}{\gamma^2}$ for different PA gel samples at various monomer (\textit{viz.,} 7.5\% and 15\% acrylamide, w/v), and cross-link (between 0.005\% and 0.12\% bisacrylamide, w/v) concentrations. Because different PA gel samples have different number of tracer beads dispersed in them, this alternative approach is pursued where every tracer bead across different samples is weighed equally. In this method, the mean square non-affine displacement collected from various samples at a given acrylamide and bis concentration is plotted against the mean square fitted strain, where all the fitted strain values are very close to the externally applied strains. The linear fit of the mean square non-affine displacements versus the mean square fitted strains, for different externally applied strains, gives the ensemble averaged $\frac{\mathcal{A}}{\gamma^2}$, as shown in Eq.~(\ref{EQ:ensemble}).  Error bars in the figure reflect the error in the linear fits. Note that the strain-normalized non-affinity parameters obtained from this method are very similar to that shown in Fig. 4b in the main paper, and confirm the robustness of our results.

\begin{eqnarray}\label{EQ:ensemble}
\mathcal{A}= \frac{1}{\sum_{i} N_i}\sum^{m}_{i=1} \sum^{N_j}_{j=1} | \vec{u}_{ij} |^2, \nonumber
\end{eqnarray}
\begin{eqnarray}
\gamma^2= \frac{1}{m}\sum^{m}_{i=1}\gamma_i^2	.
\end{eqnarray}
Here $i=1,2,\ldots,m$ labels the PA gel samples at a given acrylamide and bis concentration, with the $i$-th sample contains $N_i$ beads labeled by $j=1,2,\ldots,N_i$.

$\frac{\mathcal{A}}{\gamma^2}$ calculated from individual samples at various acrylamide and bis concentrations are listed in Table~\ref{TAB:List}, along with their respective error estimates.

\newpage

\section{for Table of Contents use only}

\begin{figure}[htp]
{\includegraphics[trim = 15mm 40mm 15mm 35mm, clip, width=0.95\textwidth]{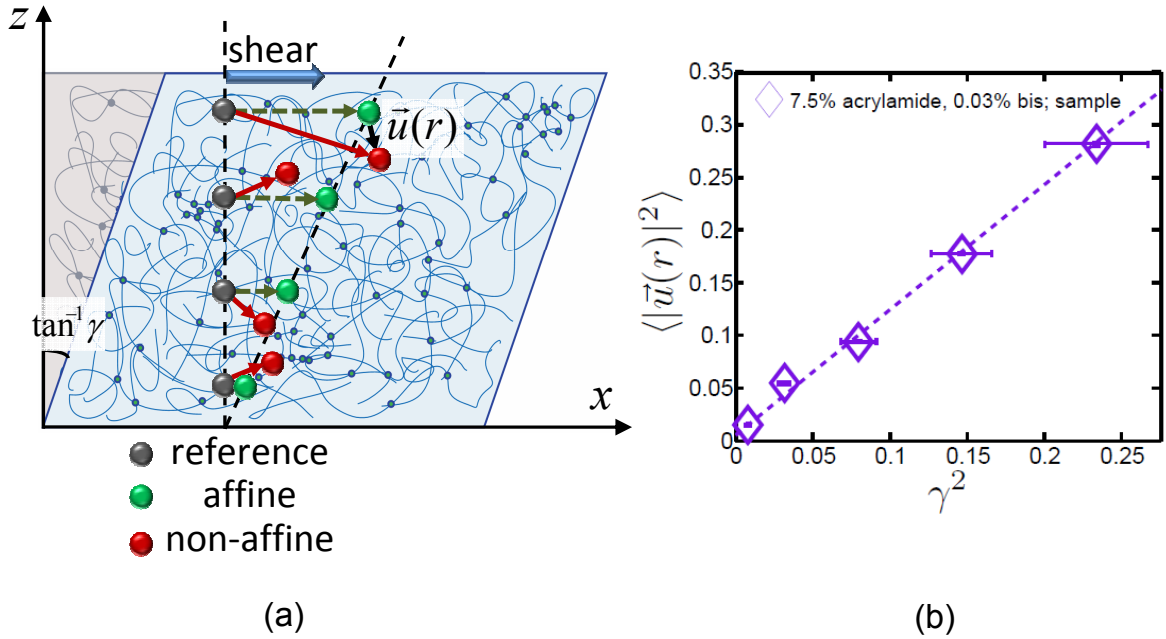}}
\caption{(a) Schematic showing shear deformation of a polymer network. Tracer beads marked in gray, green and red indicate the unsheared, affine and non-affine displacement positions under shear, respectively.  An applied strain, $\gamma$ induces a non-affine displacement, $\vec{u}(r)$, for a given tracer bead in a network. (b) Mean square non-affine displacements, $\langle |\vec{u}(r)|^2 \rangle $ of tracer beads trapped in a polyacrylamide gel scale as the square of the applied strain, $\gamma$.   }
\end{figure} 

Title: Non-affine displacements in flexible polymer networks

Authors: Anindita Basu, Qi Wen, Xiaoming Mao, T. C. Lubensky, Paul A. Janmey, and A. G. Yodh

\textit{Department of Physics and Astronomy, University of Pennsylvania,PA 19104, USA, and Institute for Medicine and Engineering, University of Pennsylvania, Philadelphia, PA 19104, USA}
 

\begin{thebibliography}{achemso}
\bibitem{1} Rubinstein, M.; Panyukov, S. \textit{Macromolecules} \textbf{1997}, \textit{30}, 8036-8044.
\bibitem{2} DiDonna, B. A.; Lubensky, T. C. \textit{Phys. Rev. E} \textbf{2005}, \textit{72}, 066619.
\bibitem{3} Rubinstein, M.; Panyukov, S. \textit{Macromolecules} \textbf{2002}, \textit{35}, 6670-6686.
\bibitem{4} Glatting, G.; Winkler, R. G.; Reineker, P. \textit{Polymer} \textbf{1997}, \textit{38}, 4049-4052.
\bibitem{5} Everaers, R. \textit{Eur. Phys. J. B} \textbf{1998}, \textit{4}, 341-350.
\bibitem{6} Svaneborg, C.; Grest, G. S.; Everaers, R. \textit{Phys. Rev. Lett.} \textbf{2004}, \textit{93}, 257801.
\bibitem{7} Sommer, J. U.; Lay, S. \textit{Macromolecules} \textbf{2002}, \textit{35}, 9832-9843.
\bibitem{8} Head, D. A.; Levine, A. J.; MacKintosh, F. C. \textit{Phys. Rev. Lett.} \textbf{2003}, \textit{91}, 108102.
\bibitem{9} Head, D. A.; MacKintosh, F. C.; Levine, A. J. \textit{Phys. Rev. E} \textbf{2003}, \textit{68}, 025101(R).
\bibitem{10} Wilhelm, J.; Frey, E. \textit{Phys. Rev. Lett.} \textbf{2003}, \textit{91}, 108103.
\bibitem{11} Mackintosh, F. C.; Kas, J.; Janmey, P. A. \textit{Phys. Rev. Lett.} \textbf{1995}, \textit{75}, 4425-4428. 
\bibitem{12} Tanguy, A.; Wittmer, J. P.; Leonforte, F.; Barrat, J. L. \textbf{2002}, \textit{Phys. Rev. B} \textit{66}, 174205.
\bibitem{13} Langer, S. A.; Liu, A. J. \textit{J. Phys. Chem. B} \textbf{1997}, \textit{101}, 8667-8671.
\bibitem{14} Rubinstein, M.; Colby, R. \textit{Polymer Physics}; Oxford University Press: New York, 2003.
\bibitem{15} Wen, Q.; Basu, A.; Winer, J.; Yodh, A.; Janmey, P. \textit{New J. Phys.} \textbf{2007}, \textit{9} 428.
\bibitem{16} Treloar, L. R. G. \textit{The Physics of Rubber Elasticity}; Clarendon: Oxford, 1975.
\bibitem{17} Crocker J.; Grier, D. \textit{J. Colloid Interface Sci.} \textbf{1996}, \textit{179} 298-310.
\bibitem{18} Gao, Y.; Kilfoil, M. L. \textit{Optics Express} \textbf{2009}, \textit(17) 4685-4704.
\bibitem{19} Mao, X.; Goldbart, P. M.; Xing, X.; Zippelius, A. \textit{Phys. Rev. E} \textbf{2009}, \textit{80}, 031140.
\bibitem{20} Liu, J.; Koenderink, G.; Kasza, K.; MacKintosh, F.; Weitz, D. \textit{Phys. Rev. Lett.} \textbf{2007}, \textit{98}, 198304.
\bibitem{21} Huisman, E. M.; Storm, C.; Barkema, G. T. \textit{Phys. Rev. E} \textbf{2008}, \textit{78}, 051801.
\bibitem{22} de Gennes, P. G. \textit{Scaling Concepts in Polymer Physics}; Cornell University Press: Ithaca, NY, 1979.
\bibitem{23} Hecht, A.; Duplessix, R.; Geissler, E. \textit{Macromolecules} \textbf{1985}, \textit{18}, 2167-2173.
\bibitem{24} Mallam, S.; Horkay, F.; Hecht, A.; Geissler, E. \textit{Macromolecules} \textbf{1989}, \textit{22}, 3356-3361.
\bibitem{25} Yeung, T.; Georges, P. C.; Flanagan, L. A.; Marg, B.; Ortiz, M.; Funaki, M.; Zahir, N.; Ming, W.; Weaver, V.; Janmey, P. A. \textit{Cell Motil. Cytoskeleton} \textbf{2005}, \textit{60}, 24-34.
\bibitem{26} Cohen, Y.; Ramon, O.; Kopelman, I. J.; Mizrahi, S. \textit{J. Polym. Sci., Polym. Phys. Ed}., \textbf{1992}, \textit{30}, 1055-1067.
\bibitem{27} Bastide, J.; Mendes Jr., E.; Boue, F.; Buzier, M.; Linder, P. \textit{Makromol. Chem., Macromol. Symp.} \textbf{1990}, \textit{40}, 81-99.
\bibitem{28} Richards, E. G.; Temple, C. J. \textit{Nature (Physical Science)} \textbf{1971}, \textit{230}, 92-96.
\bibitem{29} Hsu, T.; Cohen, C. \textit{Polymer} \textbf{1984}, \textit{25}, 1419-1423.
\bibitem{30} Hsu, T.; Ma, D. S.; Cohen, C. \textit{Polymer} \textbf{1983}, \textit{24}, 1273-1278.
\bibitem{31} Lopatin, V. V.; Askadskii, A. A.; Peregudov, A. S.; Vasil'ev, V. G. \textit{J. App. Polym. Sci.} \textbf{2005}, \textit{96}, 1043-1058.
\bibitem{32} Weiss, N.; Silberberg, A. \textit{British Polym. J.} \textbf{1977}, \textit{9}, 2, 144-150.
\bibitem{33} Weiss, N.; Vliet, T. V.; Silberberg, A. \textit{J. Polym. Sci.: Polymer Physics Edition}, \textbf{1979}, \textit{17}, 2229-2240.
\bibitem{34} Pekcan, O.; Kara, S. \textit{Polym. Int.} \textbf{2003}, \textit{52}, 676-684.
\bibitem{35} Lindemann, B.; Schroder, U. P.; Oppermann, W. \textit{Macromolecules} \textbf{1997}, \textit{30}, 4073-4077.
\bibitem{36} Baselga, J.; HernPndez-Fuentes, I.; PiBrola, I. F.; Llorente, M. A. \textit{Macromolecules} \textbf{1987}, \textit{20}, 3060-3065.
\bibitem{37} Nossal, R. \textit{Macromolecules} \textbf{1985}, \textit{18} 49-54.
\bibitem{38} Kizilay, M. Y.; Okay, O. \textit{Macromolecules} \textbf{2003}, \textit{36}, 6856-6862.
\bibitem{39} Solubility of bis in water = 0.01-0.1 g/100 ml at 18$^o$C. Solubility of acrylamide in water = 216 g/100 ml.
\bibitem{40} Wen, Q.; Janmey, P. (manuscript in preparation): \textit{An Atomic Force Microscope (DAFM, Veeco, Woodbury, NY) with a sharp conical tip is used to perform nanoindentation on a polyacrylamide gel made of 7.5\% acrylamide and 0.1\% bis. Working at the ``force volume" mode, the AFM scans an area of $1~\mu m \times 1~\mu m$ area with a resolution of 16 pixels/$\mu$m. At each pixel, a force-indentation curve is obtained and fit to the Hertz model to get the local Young's modulus. Thus, a map of Young's moduli is obtained for a $1~\mu m^2$ area with a spacial resolution of $62.5~nm$. The Young's moduli within the map varies from $3800~Pa$ to $6300~Pa$ with a mean of $4800~Pa$. The length scale of inhomogeneity is approximately $200~nm \pm 100~nm$.}
\bibitem{41} Chen, D. T.; Weeks, E. R.; Crocker, J. C.; Islam, M. F.; Verma, R.; Gruber, J.; Levine, A. J.; Lubensky, T. C.; Yodh, A. G. \textit{Phys. Rev. Lett.} \textbf{2003}, \textit{90}, 108301.
\bibitem{42} Starrs, L.; Bartlett, P. \textit{Faraday Discuss.} \textbf{2003}, \textit{123}, 323-334.
\end{thebibliography}
\end{document}